 \documentclass[%
  reprint,
superscriptaddress,
  amsmath,amssymb,
  aps,
  prb,
 floatfix,
 ]{revtex4-1}

\usepackage{graphicx,graphics}
\usepackage{dcolumn}
\usepackage{bm}
\usepackage{gensymb} 

\usepackage{bibunits}

\newcommand{\ket}[1]{\lvert #1\rangle}
\newcommand{\bra}[1]{\langle#1 \rvert}
\newcommand{\abs}[1]{\lvert #1 \rvert}

\newcommand{\braket}[2]{\langle #1 \rvert #2\rangle}

\newcommand{\br}{\mathbf{r}}
\newcommand{\bq}{\mathbf{q}}
\newcommand{\bk}{\mathbf{k}}
\newcommand{\bK}{\mathbf{K}}

\begin{document}

\preprint{APS/123-QED}

\title{Weak localization measurements of electronic scattering rates in Li-doped epitaxial graphene}

\author{A. Khademi}
	\affiliation{Stewart Blusson Quantum Matter Institute, University of British Columbia, Vancouver, BC, V6T1Z4, Canada}
	\affiliation{Department of Physics and Astronomy, University of British Columbia, Vancouver, BC, V6T1Z1, Canada}
     \affiliation{Present address: Department of Electrical and Computer Engineering, University of Victoria, Victoria, British Columbia V8P 5C2, Canada}

\author{K. Kaasbjerg}
\affiliation{Center for Nanostructured Graphene (CNG), Department of Physics, Technical University of Denmark, DK-2800 Kongens Lyngby, Denmark}

\author{P. Dosanjh}
	\affiliation{Stewart Blusson Quantum Matter Institute, University of British Columbia, Vancouver, BC, V6T1Z4, Canada}
	\affiliation{Department of Physics and Astronomy, University of British Columbia, Vancouver, BC, V6T1Z1, Canada}

\author{A. St{\"o}hr}
	\affiliation{Max Planck Institute for Solid State Research, 70569 Stuttgart, Germany}
\author{S. Forti}
	\affiliation{Max Planck Institute for Solid State Research, 70569 Stuttgart, Germany}
    \affiliation{Present address: Centre for Nanotechnology Innovation IIT@NEST, Piazza San Silvestro 12, 56127 Pisa, Italy}
\author{U. Starke}
	\affiliation{Max Planck Institute for Solid State Research, 70569 Stuttgart, Germany}

\author{J. A. Folk}%
\email{jfolk@physics.ubc.ca}
	\affiliation{Stewart Blusson Quantum Matter Institute, University of British Columbia, Vancouver, BC, V6T1Z4, Canada}
	\affiliation{Department of Physics and Astronomy, University of British Columbia, Vancouver, BC, V6T1Z1, Canada}	




\date{\today}

\begin{abstract}
Early experiments on alkali-doped graphene demonstrated that the dopant adatoms modify the conductivity of graphene significantly, as extra carriers enhance conductivity while Coulomb scattering off the adatoms suppresses it. However, conductivity probes the overall scattering rate, so a dominant channel associated with long-range Coulomb scattering will mask weaker short-range channels.  We present weak localization measurements of epitaxial graphene with lithium adatoms that separately quantify intra- and intervalley scattering rates, then compare the measurements to tight-binding calculations of expected rates for this system.  The intravalley rate is strongly enhanced by Li deposition, consistent with Coulomb scattering off the Li adatoms.  A simultaneous enhancement of intervalley scattering is partially explained by extra carriers in the graphene interacting with residual disorder.  But differences between measured and calculated rates at high Li coverage may indicate adatom-induced modifications to the band structure that go beyond the applied model.  Similar adatom-induced modifications of the graphene bands have recently been observed in ARPES, but a full theoretical understanding of these effects is still in development.
\end{abstract}


\pacs{Valid PACS appear here}
\maketitle



\begin{bibunit}[apsrev4-1] 
Adatoms have frequently been proposed as a way to alter the electronic properties of graphene: to make it superconducting,\cite{LiSC.NatPhys2012,SC.Alkali.honeycomb.PhysRevB.2015, Plasmon.SC.PhysRevLett.2007, Chiral.SC.Nat.Phys2012} magnetic,\cite{Flourine.PhysRevLett2012, 3dTransition.PhysRevLett.2013} or even a topological insulator.\cite{Franz.PhysRevX.1,Franz.PhysRevLett.2012}  Despite the conceptual simplicity of depositing selected elements onto the exposed surface of a graphene sheet, many of the more exotic predictions for novel adatom-induced electronic states in graphene have proven difficult to realize in experiment.  In order to push this area forward, experimental feedback is needed to clarify the impact of adatoms on the electronic properties of graphene.

The interaction of alkali adatoms with graphene is expected to be particularly simple, and represents a logical starting point to address the graphene-adatom puzzle. Alkali atoms are known to be efficient dopants, transferring around one electron each to the graphene lattice\cite{K.Nat.Phys.2008,DFT.Metals.PhysRevB.2008, Bonding.Metal.adatom.graphene.PhysRevB.2011} while the positively-charged ions that remain cause strong Coulomb scattering.\cite{K.Nat.Phys.2008, Ca.SolidStateCommunications2012,K.PhysRevLett.2011} The graphene-lithium system is especially interesting due to a recent report of superconductivity with a critical temperature near 6 K.\cite{Bart}  More generally, a variety of recent results indicate that adatoms must be thought of as fundamentally modifying the graphene band structure rather than than simply as perturbations on the conventional Dirac structure.\cite{Rotenberg.Extended, Gruneis.Observation, Bart,Kristen}

\begin{figure}[t]
  \centering
  \includegraphics{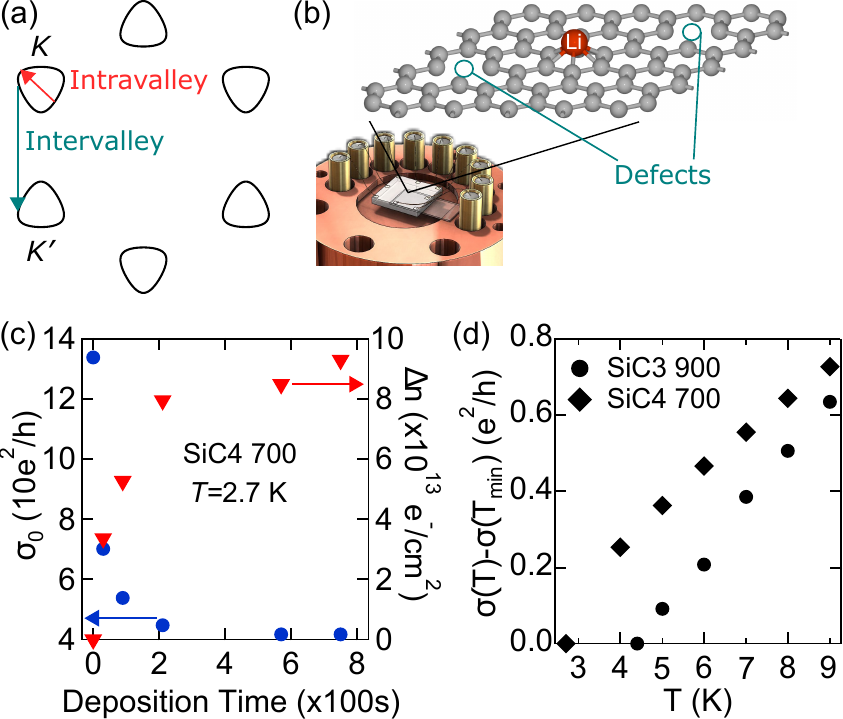}
  \caption{\small (a)  Intra- and intervalley scattering processes illustrated in a constant energy contour.  (b) Annealing stage, showing the SiC chip glued to end of a quartz plate. Illustration of graphene lattice on the chip, with vacancies that would cause intervalley scattering as well as a Li adatom. (c) Decrease in conductivity due to consecutive Li depositions. Right axis: Li-induced carrier density $\Delta n \equiv n-n_0$, starting from initial density $n_0=2.18\times 10^{13}$~cm$^{-2}$. (d)  Conductivity of graphene samples decreases monotonically with temperature down to the lowest temperatures ($T_{\rm min}$) accessed in our measurements, even after depositing Li to the point where the carrier density saturated. This panel shows the temperature-dependent conductivity change compared to the conductivities at $T_{\rm min}$: $\sigma(T_{\rm min}=2.7~{\rm K})=41.8$ e$^2$/h for SiC4 700 and $\sigma(T_{\rm min}=4.4~{\rm K})=35.1$ e$^2$/h for SiC3 900.}
  \label{Fig1}
\end{figure}

Here, we present magnetoresistance measurements of weak localization (WL) in Li-doped graphene that  probe the interaction between graphene's conduction electrons and the Li adatoms.  The analysis of WL data offers detailed information about intra- and intervalley scattering channels, which are depicted schematically in Fig.~\ref{Fig1}(a). In addition to the expected enhancement of intravalley scattering, our data indicate that intervalley scattering between graphene's $K$ and $K'$ valleys is strongly enhanced at high Li coverage.  The increase of intervalley rate due to alkali adatoms is reminiscent of a previous report in Li-intercalated bilayer graphene.\cite{NNano}

At first glance these results are surprising, because scattering off Li is expected to be long-range in character, and therefore not capable of inducing the large momentum shifts required for intervalley scattering 
[Fig.~\ref{Fig1}(a)].  In this way, lithium contrasts with other adatoms and substitutionals that are expected to introduce both Coulomb and short-range scattering in graphene.\cite{Bart.Tl.Graphene.long.and.short.range.scatt.Nano.Lett.2015,nitrogen.graphene.intervalley.short.range.scattering.acsnano.2017,fluorinated.graphene.theory.exp.PhysRevB.2019}  Our data can partially be accounted for through enhanced scattering off pre-existing short-range disorder, as confirmed by a tight binding analysis of scattering rates and conductivity that includes trigonal warping and the nonlinearity in the band structure away from the Dirac point. But a discrepancy remains between experimental data and tight-binding predictions for the intervalley rate at high Li coverage, pointing to  adatom-induced bandstructure modifications that go beyond our modelling.  Such modifications would be consistent with ARPES experiments\cite{Rotenberg.Extended, Gruneis.Observation, Bart} and recent theoretical calculations.\cite{Kristen}

Measurements are reported on four epitaxial monolayer graphene samples: SiC1 was grown on a weakly-doped 6H-SiC(0001) surface;\cite{SiC.PhysRevB.84.125449} SiC2-4 were cut from commercially available epitaxial graphene grown on the semi-insulating 4H-SiC(0001) surface.\cite{graphensic} The labelling of SiC1-4 is consistent with an earlier doping study on these samples,\cite{My.Li.Doping.Paper} where further sample details can be found.  After growth, eight contacts were deposited by thermal evaporation onto the corners and edges of each sample, using shadow evaporation to avoid polymer resist contamination.  Resistances were measured in a 4-probe quasi-van der Paaw configuration, then converted to conductivities for comparison with weak localization theory.

Experiments were performed in a UHV chamber with base pressure below $5\times$10$^{-10}$ torr, with Li evaporated from an SAES getter source while the sample was held at 4 K on a liquid-He cooled cold finger. A custom stage [Fig.~\ref{Fig1}(b)] enabled annealing operations up to 900 K while also ensuring cryogenic thermal contact between the sample and the cold finger during transport measurements.\cite{My.Li.Doping.Paper}  The stage could be cooled below 3~K by pumping on the liquid He line. Photographs of several samples on this stage can be seen in supplemental Fig.~S1.\cite{supplemental}

The first step in each experiment was a 3-day bakeout of both sample and chamber at 390 K. For some samples, further annealing of the chip was performed using the stage [Fig.~\ref{Fig1}(b)].\cite{My.Li.Doping.Paper} Then, the sample and a surrounding shroud were cooled down to 3-4 K, and Li was deposited in multiple increments.  The shroud was open only during Li depositions, then closed again before magnetoresistance measurements were performed.  Carrier density was determined by transverse magnetoresistance (the classical Hall effect) after each deposition, while the scattering rates that are central to this paper were determined from the longitudinal magnetoresistance through WL.

It has previously been shown that high temperature annealing prior to Li deposition is crucial to achieving efficient graphene-Li coupling.\cite{My.Li.Doping.Paper}  Here, we explore samples with a range of preparations: SiC1 and SiC2 were measured with no higher temperature anneals following the 390 K bakeout. SiC3 underwent one Li deposition-and-measurement sequence right after bakeout, then it was annealed at 900 K (which desorbed the Li) and a second Li deposition-measurement sequence was performed.  SiC4 was annealed first at 500 K, then a Li deposition-measurement sequence was performed, then it was annealed again at 700 K before a second deposition-measurement sequence. For clarity, data from a given sequence is labelled by the sample name and the most recent annealing temperature in Kelvin.  For example, SiC1 390 refers to sample SiC1 with no additional anneal after the 390 K bakeout.

Figure \ref{Fig1}(c) illustrates an example of doping level and conductivity changes resulting from consecutive Li depositions. For SiC4 700, the induced carrier density due to Li saturated around $10^{14}$ e$^-$/cm$^2$ while the conductivity decreased by a factor of four.  For SiC3 900, annealed at a higher temperature, the saturation carrier density was a factor of two larger [Fig.~S2(a) \cite{supplemental}]. The saturation of carrier density in our samples, with increasing Li deposition, was discussed in Ref.~\onlinecite{My.Li.Doping.Paper}, and presumably results from insufficient surface preparation.  

All samples showed a weakly insulating temperature dependence of conductivity below around 10 K. Fig.~\ref{Fig1}(d) shows this behaviour for SiC3 900 and SiC4 700 after their final Li depositions; see supplemental Fig.~S2(b) for SiC3 390 and SiC4 500 \cite{supplemental}.  The observed conductivities were consistent in all cases with the logarithmic dependence expected for weak localization and the electron-electron correction to conductivity in 2D.  The fact that the conductivity changed smoothly with the cold finger temperature down to 2.7 K confirms the efficient thermal coupling of our sample stage design.  No upturn in conductivity at low temperature was observed in any samples, as might have been expected if superconductivity ($T_c\sim 6$ K) were induced in these samples by the Li.\cite{Bart}

\begin{figure}[t]
  \centering
  \includegraphics{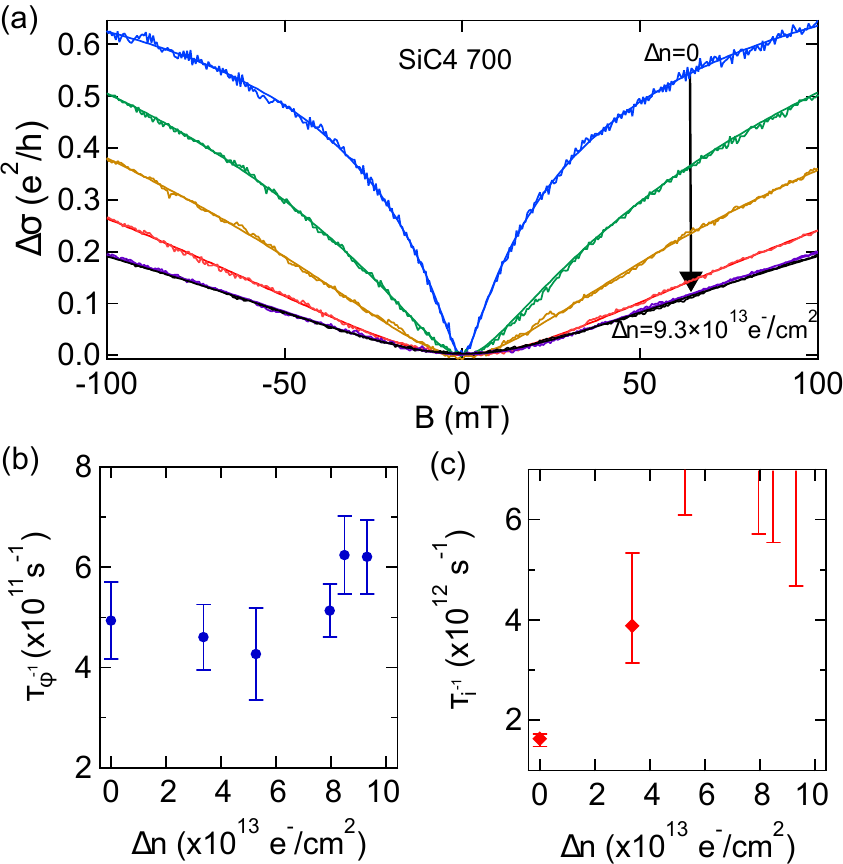}
  \caption{\small (a) The effect of Li deposition on magnetoconductivity, $\Delta \sigma \equiv \sigma(B)-\sigma(B=0)$.  As in Fig.~\ref{Fig1}, $\Delta n$ represents increase carrier density due to Li, starting from  $n_0=2.18\times10^{13}$ cm$^2$. The solid lines are fits to Eq.~\ref{eq1}. 
  Extracted dephasing (b) and  intervalley (c) rate versus induced carrier density due to Li.  All data correspond to SiC4 700 at $T=$2.7 K.}
  \label{Fig2}
\end{figure} 

The expected WL dip in longitudinal conductivity at zero magnetic field [Fig. \ref{Fig2}(a)] was observed in all samples.  Electronic scattering rates were extracted by fitting to the standard WL form for graphene:\cite{McCann.PhysRevLett.2006}
\begin{align}
\label{eq1}
	\Delta \sigma (B_{\perp}) & = \sigma(B_\perp)-\sigma(0) 
    	= \frac{e^2}{\pi h}\left [F \left(\tfrac{\tau^{-1}_B}{\tau^{-1}_\varphi}\right)  \right.
    \nonumber \\
    & \quad 
	\left. 
    	- F\left(\tfrac{\tau^{-1}_B}{\tau^{-1}_{\varphi}+2\tau^{-1}_\text{i}} \right)
    	-2F\left( \tfrac{\tau^{-1}_B}{\tau^{-1}_{\varphi}+\tau^{-1}_{\ast}+\tau^{-1}_\text{i}} \right)
    	\right]
\end{align}
where $F(z)=\ln(z)+\psi(\frac{1}{z}+\frac{1}{2})$, $\psi$ is the digamma function and $\tau^{-1}_B=4eDB_\perp/\hbar$ is the phase accumulation rate in magnetic field $B_\perp$ with diffusion constant $D$.   $\tau_\varphi^{-1}$ represents the conventional phase decoherence rate known from WL studies in metals. $\tau_\text{i}^{-1}$ and $\tau_\ast^{-1}$ are the intervalley  and intravalley scattering rates corresponding to scattering between or within a single valley, respectively [Fig.~\ref{Fig1}(a)].  $\tau_\ast^{-1}$ is very high in epitaxial graphene, even without Li, due to  chirality-breaking disorder and trigonal warping.\cite{McCann.PhysRevLett.2006,intervalley.J.Phys.:Condens.Matter2010,wl_gorbachev}  As a result, the last term in Eq.~\ref{eq1} is suppressed and not included in our fits.

Extracted values of $\tau^{-1}_{\varphi}$  were nearly independent of Li coverage, even over an order of magnitude increase in carrier density [Fig.~\ref{Fig2}(b)]. This can be understood from the fact that Li is a light adatom, and not a source of spin-orbit coupling or magnetism~\cite{Franz.PhysRevX.1}. The contribution to the dephasing rate due to electron-electron interactions would be expected to rise from 11 ns$^{-1}$ to 26 ns$^{-1}$ for the data in Fig.~2, as conductivity decreased from 134 to 42$e^2/h$ with added Li [Fig. \ref{Fig1}(c)].\cite{wl_gorbachev, supplemental}  However, this represents a small perturbation on the overall dephasing rate, which, in epitaxial graphene on SiC, is dominated by magnetic impurities.\cite{PhysRevLett.115.106602, PhysRevLett.107.166602}

\begin{figure}[t]
  \centering
  \includegraphics{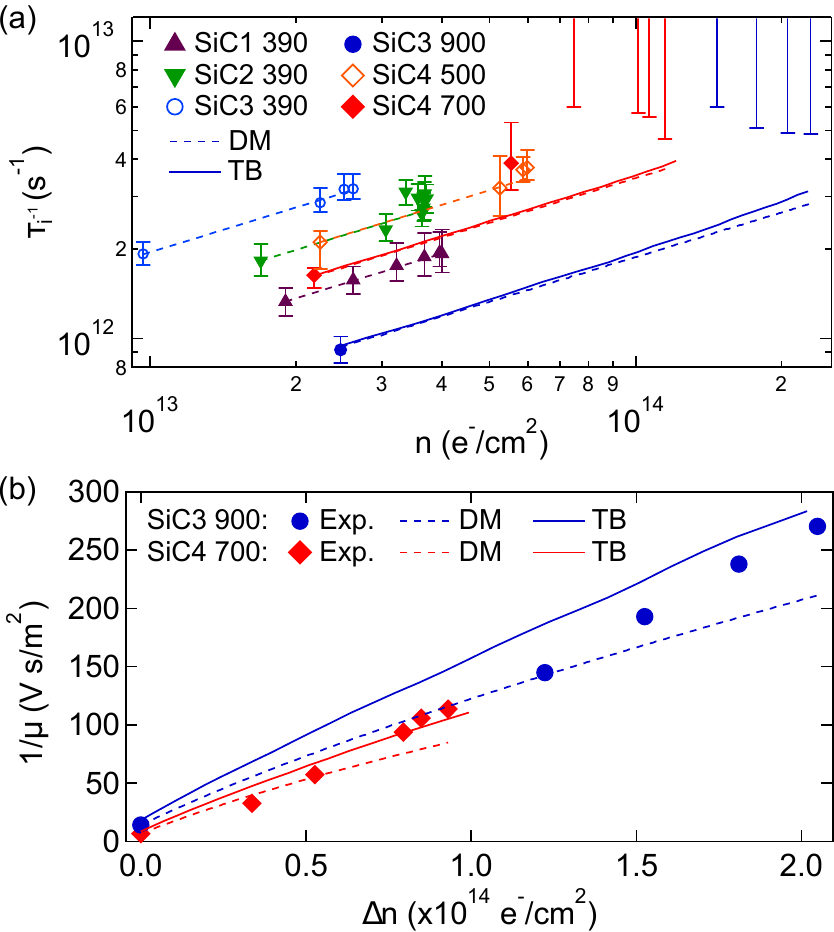}
  \caption{\small (a) Intervalley rates for SiC1-4 through multiple sequences of Li deposition, shown in log-log scale to highlight the power-law behaviour.  (b) The inverse mobility versus change of charge carrier density induced by Li deposition for SiC3 and SiC4, which were annealed to 700 and 900 K prior to cryogenic Li deposition. The dashed and solid lines in  Figs. 3(a) and 3(b) show theoretical predictions based on the Dirac model (DM) and a tight-binding (TB) description, respectively.
  }
  \label{Fig3}
\end{figure}

In contrast, $\tau^{-1}_\text{i}$ increased significantly after Li deposition [Fig.~\ref{Fig2}(c)], ultimately to values so high that the second term in Eq.~\ref{eq1} was suppressed and the error bars in the extracted $\tau^{-1}_\text{i}$ extend off the top of the graph [see Ref.~\onlinecite{supplemental} for details on fitting]. These half-error-bars indicate that the extracted $\tau_i$ was indistinguishable from zero within experimental uncertainty, which was limited primarily by the 100 mT scan range of the coil.

Figure \ref{Fig3}(a) compiles $\tau^{-1}_\text{i}$ for 6 samples, presenting a series of Li depositions for each sample.  It confirms the consistently strong increase of intervalley scattering as Li is added, in spite of the common expectation that alkali adatoms should have minimal effect on intervalley scattering.\cite{K.Nat.Phys.2008,K.PhysRevLett.2011,Intervalley.charge.state.of.defects.PhysRevB.2016}  A clue to understanding this surprising result comes from the functional form of the scattering rate increase, seen clearly in the log-log plot of Fig.~\ref{Fig3}(a): the measured $\tau^{-1}_i$ fits well to a $\tau^{-1}_i\propto \sqrt{n}$ dependence (dashed lines) up to a carrier density around $5\times 10^{13}$ cm$^{-2}$.  Scattering rates for a given density of short-range  scatterers would generically be proportional to the graphene density of states, which is $D(E_F) = 2\sqrt{n}/(\sqrt{\pi} \hbar v_F)$ within the linear Dirac model for graphene's band structure ($E=\hbar v_F k$).   Thus, a $\sqrt{n}$ dependence is expected purely due to the doping effect from Li, enhancing the scattering rate from pre-existing short-range defects in graphene on SiC\cite{SR.disorder.SiC.graphene.Phys.Rev.B.2012} via the graphene density of states.

With $\tau_i^{-1}$  extracted from WL, $\tau_*^{-1}$ can be then be determined  from mobility as described in the supplement [Eq.~S21] \cite{supplemental}. Figure~\ref{Fig3}(b) illustrates the inverse mobility, $\mu^{-1}=en/\sigma$, for the two samples with highest carrier density.  The close-to-linear relationship between $\mu^{-1}$ and $\Delta n$ can also be explained within the Dirac model.  In our experiment, the change in graphene carrier density, $\Delta n$, is proportional to the density of Li adatoms, $n_\text{Li}$.  When conductivity is limited by Coulomb scattering off charged Li,\cite{K.Nat.Phys.2008,Ca.SolidStateCommunications2012,K.PhysRevLett.2011} one expects $\sigma\sim n / n_\text{Li}$ giving $\mu^{-1} = (\sigma / en)^{-1} \sim \Delta n$.  

The discussion above demonstrates that the modifications to intra- and intervalley scattering rates for low levels of Li doping can be approximately explained by the linear Dirac model (DM). Above $5\times 10^{13}$ cm$^{-2}$, however, the intervalley data in Fig.~\ref{Fig3}(a) lies well above the $\sqrt{n}$ traces on the graph,  indicating either (i) new short-range scatterers  being added or activated, and/or (ii) deviations from the linear Dirac-cone density of states.  The fact that the divergence between intervalley data and calculations only appears at high doping levels, and that Li adatoms or clusters would not be expected to bond strongly enough with the graphene to act as short-range scatterers, \cite{DFT.Metals.PhysRevB.2008,Bonding.Metal.adatom.graphene.PhysRevB.2011,
Li.Cluster.ACS.Appl.Mater.Interfaces.2013,Li.Cluster.J.Phys.Chem.Lett.2014} indicates that option (i) is unlikely.

In order to evaluate the second option, we  perform numerical calculations of the scattering rate and conductivity/mobility based on the nearest-neighbor tight-binding (TB) description of the graphene bands. The TB description accounts for trigonal warping of the Dirac cones, illustrated by the constant energy contours in Fig.~\ref{Fig1}(a), as well as nonlinear corrections to the Dirac model.  These corrections are important at the high carrier densities accessed in this work, where Fermi energies in excess of $1$ eV are achieved (for a detailed discussion of DM and TB models, see Ref.~\onlinecite{supplemental}).

Our TB analysis is compared with experimental data through a calculation of scattering rates due to randomly distributed short-range defects and Li adatoms:
\begin{equation}
\label{eq:tau_inv}
\tau_{\alpha}^{-1}(\varepsilon_\mathbf{k})
     =\frac{2\pi}{\hbar} n_\alpha \int \!\frac{d\mathbf{k}'}{(2\pi)^2} 
        \vert V_{\mathbf{k}\mathbf{k}'}^{\alpha} \vert^2 
       \delta(\varepsilon_\mathbf{k}-\varepsilon_{\mathbf{k}'}) ,
\end{equation}       
where the index $\alpha=\{\text{Li},\,\text{res}\}$ represents the disorder type, identifying whether the scattering originates from Li adatoms or from residual disorder, 
$n_\alpha$ is the areal density of the respective disorder, $\varepsilon_\bk$ is the TB band energy, and $V_{\mathbf{k}\mathbf{k}'}^{\alpha}$ is the impurity matrix element for scattering from $\mathbf{k}$ to $\mathbf{k}'$. \cite{supplemental}

Coulomb scattering by the Li adatoms is described by a matrix element $V_{\mathbf{k}\mathbf{k}'}^{\text{Li}}\propto V_C(q,d)$ that is proportional to the 2D Fourier tranform of the screened Coulomb potential, $V_C(q,d)=\tfrac{Z_\text{Li}e^2}{2\epsilon_0 \kappa \varepsilon(q)}\tfrac{e^{-qd}}{q}$.  Here  $q=\abs{\bk-\bk'}$ is the scattering vector, $\kappa=(\varepsilon_\text{SiC} +1)/2$ is the dielectric constant of the environment, $\varepsilon(q)$ is the static dielectric function of graphene, $Z_\text{Li}=0.9$ is the expected valence of Li adatoms,\cite{DFT.Metals.PhysRevB.2008,Bonding.Metal.adatom.graphene.PhysRevB.2011,Kristen} and $d=1.78$~{\AA} is the expected distance between the Li adatoms and the graphene plane.\cite{DFT.J.Phys.Chem.B2006,DFT.Li.Graphene.Gap.PhysRevB.2009,Kristen}

We assume that residual short-range disorder is dominated by atomic defects for which the scattering matrix element is momentum-independent, therefore  $V_{\mathbf{k}\mathbf{k}'}^{\text{res}} = V_{\text{res}}$, with different disorder strengths for intra- and intervalley scattering. Since $V_{\text{res}}$ is explicitly not dependent on subsequent Li deposition, its value was extracted from the initial data for each sample [see supplemental Table~I \cite{supplemental}], leaving us with no free fitting parameters in our theory.  TB and DM modelling were calculated using $n_\alpha$'s and $V$'s for the residual short-range intra- and intervalley scattering extracted from the $\Delta n=0$ values of $\mu$, and $\tau_i$ in Fig.~\ref{Fig3} (Values of $\sigma$ in Fig.~\ref{Fig1} can be used instead of $\mu$).

At low carrier densities where the DM applies, Eq.~\ref{eq:tau_inv} yields a scattering rate that scales as $\tau_{\text{res}}^{-1} =
\tfrac{n_\text{res} V_{0}^2 E_F}{\hbar^3 v_F^2} \propto \sqrt{n}$ as expected, consistent with the dependence of $\tau^{-1}_i$ below $5\times 10^{13}$ cm$^{-2}$ in Fig.~\ref{Fig3}(a).  The DM predictions (dashed lines) lie almost on top of the TB analysis (solid curves) at low density, confirming that the explanation of residual scatterers made more effective at higher carrier density survives the more accurate TB analysis.

At higher densities, the TB intervalley rates begin to deviate from the DM result due to the nonlinearity of the bands at high energies, but the effect is not nearly strong enough to account for the observed enhancement of the intervalley rate in the data. Therefore, even the second option discussed above (deviations from the linear Dirac-cone density of states) cannot explain the data within a non-interacting TB analysis.  This experimental result is, however, consistent with recent ARPES studies\cite{Rotenberg.Extended, Gruneis.Observation, Bart} and theory\cite{Kristen}, which indicate that the Dirac cone in alkali-doped graphene is strongly perturbed at high adatom densities.
It is worth noting that the match between TB modelling and experimental data is much better in the carrier mobility [Fig.~\ref{Fig3}(b)], despite the lack of free fitting parameters. This can be attributed to the fact that the conductivity is limited by intravalley Coulomb-disorder scattering, while it is only weakly dependent on residual short-range scattering. It should thus be noted that it is our combined measurement of the zero-field conductivity and WL that has permitted a detailed analysis of the individual intra- and intervalley scattering rates, and it is this analysis that confirmed the discrepancy between experimental data and TB calculations of the scattering rates. 

In summary, Li adatoms deposited in cryogenic UHV are observed to enhance both intervalley and intravalley carrier scattering rates in epitaxial graphene. The enhancement of the intravalley rates is quantitatively explained by Coulomb scattering off the ionized Li dopants that remain on the graphene surface, based on a calculation with no free fitting parameters. The enhancement of the intervalley rate, while surprising for an alkali atom like Li that bonds weakly to graphene and causes minimal short-range scattering, can largely be explained by enhanced scattering off pre-existing short-range scatters.

At the highest carrier densities observed in this work, however, deviations between our TB calculations and the experimental data do appear. This may originate from effects not accounted for by our TB model, such as the above-mentioned modifications of the graphene bands observed in ARPES and theory.\cite{Rotenberg.Extended, Gruneis.Observation, Bart,Kristen}  Other possible explanations could be:
Our TB model may use an incorrect position of the van Hove singularity in the graphene density of states, which is predicted by DFT to lie at a much lower energy\cite{Kristen}. Resonant scattering\cite{Wehling.res.PhysRevB.2009,Wehling.res.PhysRevLett.2010,Irmer.PhysRevB.2018} off the Li impurity band\cite{Bart, Kristen} may play a role, as the impurity band associated with Na ions were shown to modify the transport properties of Si MOSFETs significantly\cite{
Hartstein.Fowler.1980,
RevModPhys.1982}, but theoretical predictions for the contribution of this mechanism to intervalley scattering are too weak to explain the experimental data\cite{Kristen}. Nonlocal screening may enhance intervalley scattering by charged impurities.\cite{Intervalley.Charged.Scattering.PhysRevB.2015} Or, the Dirac cones themselves may be modified by electron-electron interactions.\cite{Schliemann:Interacting}


The data reported here present a comprehensive picture of intervalley and intravalley scattering in adatom-doped graphene.  We hope that they will help in relating ARPES and transport experiments that have until now offered disconnected pictures of scattering rates in, respectively, high and low density regimes \cite{Bart,Rotenberg.Extended, Gruneis.Observation,K.Nat.Phys.2008,Ca.SolidStateCommunications2012,K.PhysRevLett.2011,In.12K.PhysRevB.2015}.  Inconsistencies uncovered in this work point to the need for further experimental and theoretical investigation of the electronic structure and scattering mechanisms in graphene, in order to fully unravel the properties of graphene with alkali adatoms.

\section*{Acknowledgment}
The authors acknowledge D Bonn, S Burke, A Damascelli,  G Levy, B Ludbrook, A Macdonald, P Nigge, A P\'{a}lyi and E Sajadi for numerous discussions, as well as Ludbrook and J Renard for assistance in building the chamber. KK acknowledges support from the European Union's Horizon 2020 research and innovation programme under the Marie Sklodowska-Curie Grant Agreement No. 713683 (COFUNDfellowsDTU). The Center for Nanostructured Graphene (CNG) is sponsored by the Danish National Research Foundation, Project DNRF103. AK thanks UBC for financial support through the Four Year Doctoral Fellowship.  Research supported by NSERC, CFI, and the SBQMI in partnership with MPI.

\putbib[main]
\end{bibunit}
\widetext
\clearpage
\newpage
\setcounter{equation}{0}
\setcounter{figure}{0}
\setcounter{table}{0}
\setcounter{page}{1}
\renewcommand{\theequation}{S\arabic{equation}}
\renewcommand{\thefigure}{S\arabic{figure}}
\renewcommand{\bibnumfmt}[1]{[S#1]}
\renewcommand{\citenumfont}[1]{S#1}
\makeatletter
\renewcommand{\present@bibnote}[2]{}
\makeatother
\begin{bibunit}[apsrev4-1]
\setcounter{figure}{0}
\section{SUPPLEMENTARY INFORMATION}

\subsection{Photograph of SiC samples on the stage}
\label{sec:samples}

\begin{figure} [h]
  \centering
  \includegraphics[width=3.2in]{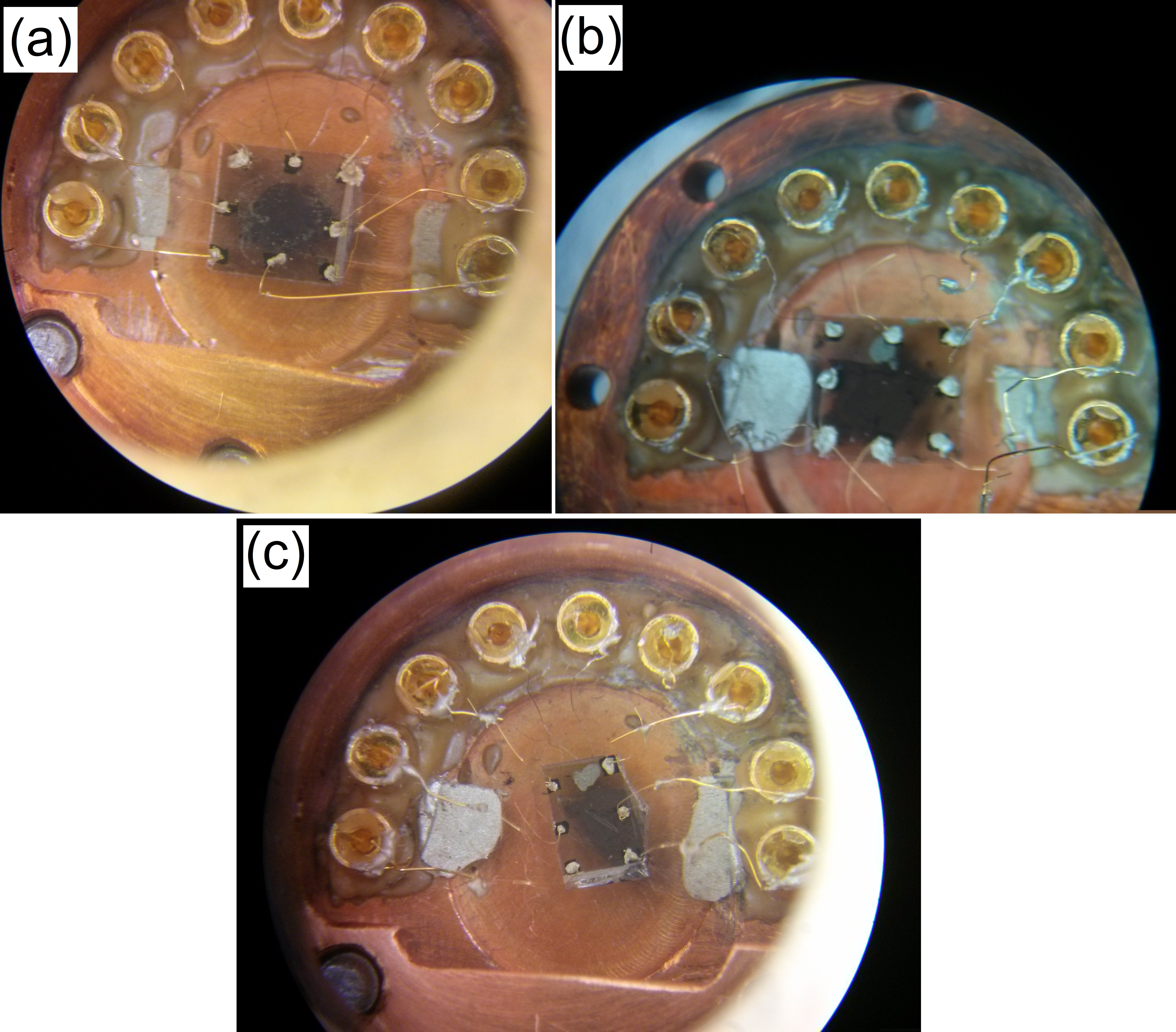}
  \caption{Photograph of different samples installed on the stage: (a) SiC2, (b) SiC3, (c) SiC4.}
  \label{fig:samplephoto}
\end{figure}

Fig.~\ref{fig:samplephoto} shows photographs of SiC2,3,4 installed on the stage. More description about the sample stage can be found in Ref.~\cite{s.My.Li.Doping.Paper}, especially in Fig. 2 from that work and discussion thereof.

\subsection{Change of conductivity by Li deposition and temperature for SiC3~900, SiC3~390, and SiC4~500}
\label{sec:intervalley times/length}

\begin{figure}[!h]
  \centering
  \includegraphics[width=0.5\textwidth]{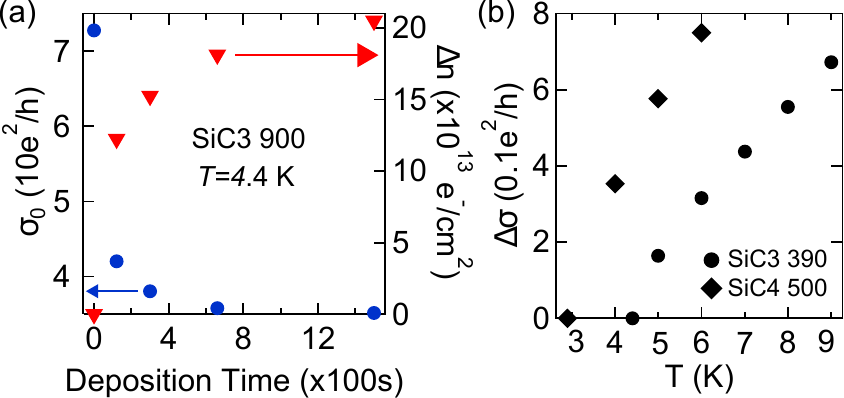}
  \caption{(a) The conductivity of SiC3 900 dropped due to Li deposition at $T$=4.4 K, as the carrier density inteased. (b) Change in conductivity $\Delta \sigma$ versus temperature $T$ after saturation of Li deposition for SiC3 390 and SiC4 500.}
  \label{FigSiC3Con}
\end{figure}

Figs.~1 (c) and (d) in the main text show how conductivity and carrier density change with deposition time for SiC4 700, and the change in conductivity with temperature $T$ after saturation of Li deposition for SiC4 700 and SiC3 900. In this section,  analogous curves for other samples are shown.  Fig.~\ref{FigSiC3Con}(a) shows the conductivity and induced carrier density (i.e. change of carrier density by Li adatoms) for SiC3 900. The induced carrier density due to Li saturated at around $2\times 10^{14}$ e$^-$/cm$^2$ while the conductivity decreased by a factor of two.  Fig.~\ref{FigSiC3Con}(b) shows the conductivity change versus temperature $T$ for SiC3 390 and SiC4 500, after depositing Li to the point that their carrier densities were saturated.  The same weakly insulating behaviour visible in the main text, Fig.~1(d), is apparent here.

\subsection{The contribution to the dephasing rate from electron-electron interactions}
\label{subsec:dephasing}

The expected contribution to the dephasing rate due to electron-electron interactions is linear in temperature \cite{s.Altshuler.PhysRevB.1980, s.PhysRevLett.115.106602}:
\begin{equation}
\label{eq:Dephasing Rate linear in T}
\tau^{-1}_\varphi=\frac{k_{B}ln(g/2)}{\hbar g}T
\end{equation}
for dimensionless conductivity $g=\sigma h/e^{2}$. For SiC4 700 at $T=$2.7 K (c.f. Fig. 2(b) in the main text), $g$ changes from 133.75 to 41.79 due to Li deposition. As a result, the calculated contribution to the dephasing rate due to electron-electron interactions changes from 11 ns$^{-1}$ to 26 ns$^{-1}$.  This is a small perturbation on the total dephasing rate from Fig. 2(b), and would not be noticeable in the data.

\subsection{Weak localization curves' fitting procedure}
\label{subsec:WL fitting procedure}

\begin{figure}[!h]
  \centering
  \includegraphics[width=0.4\textwidth]{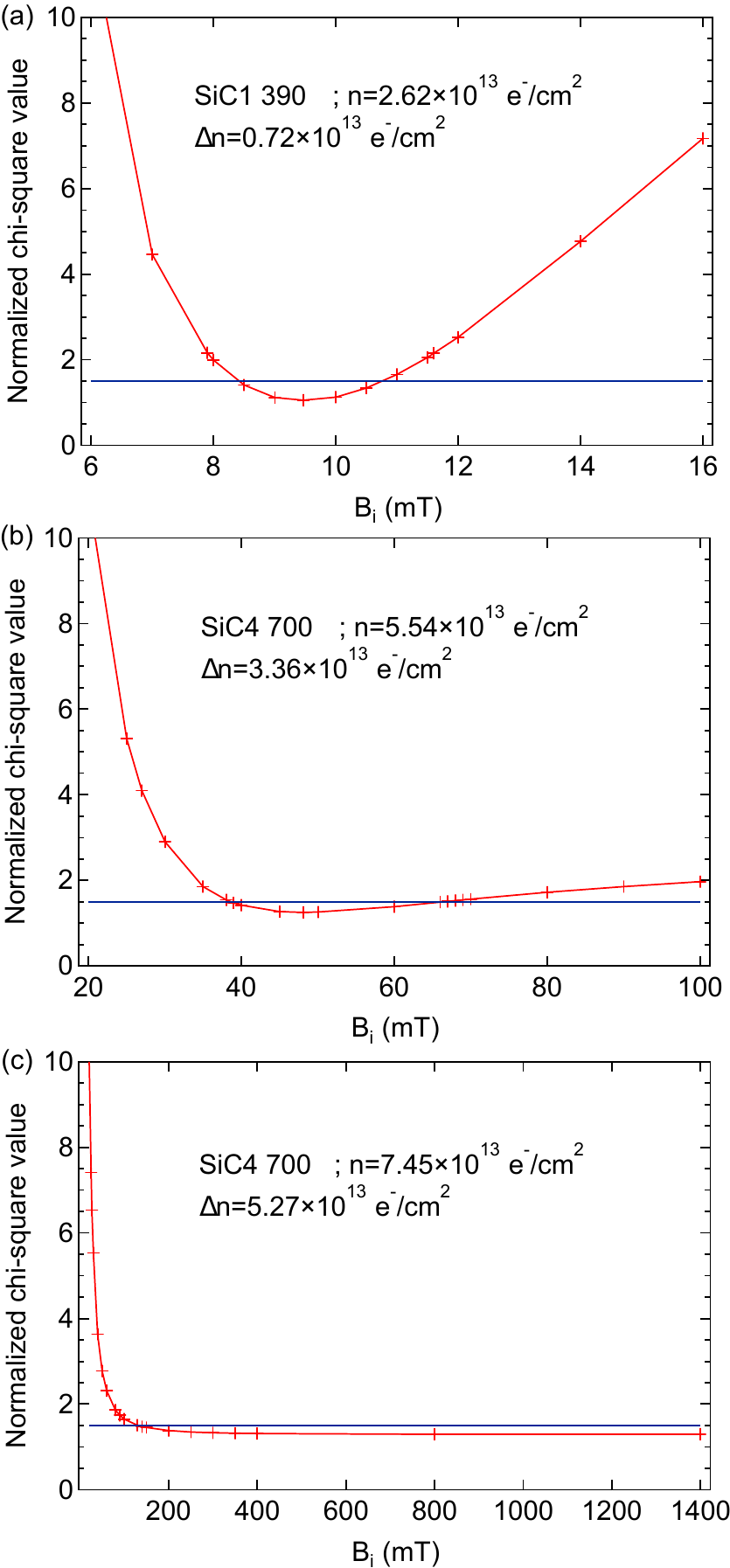}
  \caption{The chi-square value versus different possible $B_{i}$ for (a) the second point of SiC1, (b) the second point of SiC4 700, and (c) the third point of SiC4 700 in Fig.~3(a).}
  \label{figchisqr}
\end{figure}

This section describes how the range/error bars of the intervalley rates were calculated in Fig.~2(c) and Fig.~3(a) of the main text. In order to estimate the error bars for the extracted values $\tau_{i}^{-1}$, we defined the intervalley characteristic field $B_{i}=(\hbar/4eD)\tau_{i}^{-1}$, where $D$ is the diffusion constant. Then, we fit magnetoconductivity data with different $B_{i}$ and recorded the normalized chi-square value $\chi_{v}^{2}$. The chi-square may be defined as $\Sigma ((y-y_{i})/w_{i})^{2}$ where $y$ is a fitted value, $y_{i}$ is the measured data value and $w_{i}$ is the standard error for the given point. The normalized chi-square, which is also called the reduced chi-square, is defined as the chi-square per degree of freedom (i.e., number of measurements minus number of fitting parameters). While a value of $\chi_{v}^{2}$=1 indicates that the extent of the match between measurement and fit is in accord with the noise in the data, a $\chi_{v}^{2}\gg$1 indicates a poor fit \cite{s.chi.square.Book}. Figs.~\ref{figchisqr} (a) and (b) clearly show a minimum in $\chi_{v}^{2}$, $\chi_{v}^{2}\sim 1$, around $B_i=9.5$ mT and $B_i=48$ mT respectively.  These are then the best fit values.  We define the error bar in $\chi_{v}^{2}$ to be the range in $B_i$ over which $\chi_{v}^{2}<1.5$, a value that is somewhat arbitrary but not unreasonable given the trends observed over multiple datasets and multiple Li depositions seen in Fig.~3(a) in the main text. 

For some datasets, such as the highest carrier density points in Figs.~2(c) and 3(a), the fitted values of $\tau_i$ could not be distinguished from zero within experimental uncertainty.  From a practical point of view, this implies that $B_i$ was apparently above the field accessible in our hand-wound coil, 100 mT.  In the fitting process described above, this meant that $\chi_{v}^{2}$ decreased initially (starting from $B_i=0$), but then flattened out once it reached $\chi_{v}^{2}\sim 1$ and did not increase again for very high $B_i$.  A clear example is shown in Fig.~\ref{figchisqr} (c).  In this case, a lower bound for the fitted $B_i$ could be determined, as the point at which $\chi_{v}^{2}$ rose above 1.5, but there was no upper bound the error bar for $B_i$.

Because the quantitative analysis of $\chi_{v}^{2}(B_i)$ was crucial for the determination of error bars, it was important to distinguish between experimental noise, which could be safely averaged over when fitting, and real trends in the data.  As seen in the main text, the WL function changes rapidly around $B=0$, but only slowly for higher $|B|$.  To account for this, we divided  WL curves into three sections: the region around the peak and two other sections. Fifth order polynomial functions were fit to the two outer sections, $|B|>13.5$ mT, and the data was replaced with fits in those sections. The data in the central region, $|B|<13.5$ mT, was left intact.  Then, we fit the WL function [Eq.~1] to the new curve. With this method, the minimum $\chi_{v}^{2}(B_i)$ more accurately represented the fit quality and was left influenced by measurement noise. This method of fitting was used for acquiring all of the data points and error bars of annealed samples of SiC 500 K, SiC 700 K, and SiC 900 K in Fig.~3(a). For other samples, which has much lower $B_{i}$, we did not divide WL curves and fit the WL function [Eq.~1] to the whole curves.

\subsection{Theoretical tight-binding and Dirac modelling}

In this section, we describe the details of the tight-bonding (TB) and Dirac model (DM) calculations presented in the main manuscript.

Our starting point is the nearest-neighbor tight-binding model of graphene, 
\begin{equation}
    H_0 = -t \sum_{\bk} [ f_\bk c^\dagger_{A\bk} c_{B\bk}^{\phantom\dagger} +
    \mathrm{h.c.} ] \equiv \sum_{\bk} \bm{c}^\dagger_{\bk} \bm{\mathcal{H}}_\bk
    \bm{c}_{\bk}^{\phantom\dagger}     ,
\end{equation}
where $t=2.7$~eV is the hopping parameter,
\begin{equation}
    \label{eq:H}
    \bm{\mathcal{H}}_\bk = -t
    \begin{pmatrix}
        0 & f_\bk \\
        f_\bk^* & 0
    \end{pmatrix}
    , \quad
    f_\bk = 1 + e^{i\bk\cdot\mathbf{a_1}} + e^{i\bk\cdot\mathbf{a_2}} ,
\end{equation}
and $c^\dagger_{\nu\bk}$ ($c^{\phantom\dagger}_{\nu\bk}$) is the creation (annihilation)
operator for the $\nu\in\{A,B\}$ sublattice state $\ket{\nu\bk}=\tfrac{1}{\sqrt{N}}\sum_n e^{i\bk\cdot\mathbf{R}_n} \ket{\nu \mathbf{R}_n}$, $N$ is the number of unit cells, $\mathbf{R}_n= n_1 \mathbf{a}_1 + n_2 \mathbf{a}_2$ is the lattice vector to the $n$'th unit cell, and $\mathbf{a}_{1/2} = a/2 (\sqrt{3}, \pm 1)$ are the primitive lattice vectors with lattice
constant $a=2.46$~{\AA}.

\begin{figure}[!b]
  \includegraphics[width=0.5\textwidth]{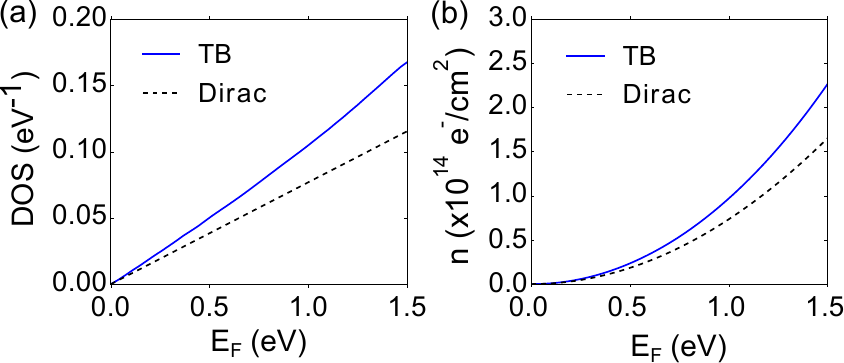}
  \caption{Comparison between the density of states (left) and carrier density vs energy $E_F$ (right) in graphene obtained with the tight-binding model (TB) and low-energy Dirac model.}
\label{fig:TB}
\end{figure}

The Bloch states $\Psi_{s\bk}(\br)$ are given by the two-component spinor eigenstates of the TB Hamiltonian in Eq.~\eqref{eq:H} as
\begin{equation}
    \label{eq:wf}
    \Psi_{s\bk}(\br) = \sum_{\nu=A,B} \chi_{\nu s\bk} \, \phi_{\nu \bk}(\br),
    \quad 
    \bm{\mathcal{\chi}}_{s\bk} = \frac{1}{\sqrt{2}}
    \begin{pmatrix}
        1 \\
        s e^{i\theta_\bk}
    \end{pmatrix} ,
\end{equation}
where $s=\pm 1$ is the band index, $\theta_\bk = \arg f_\bk$, $\phi_{\nu \bk}(\br)=\braket{\br}{\nu \bk}$, and the corresponding eigenenergies are $\varepsilon_{s\bk}=s t \abs{f_\bk}$. In Fig.~\ref{fig:TB} we show the difference in the density of states and carrier density vs $E_F$ between the tight-binding model and the Dirac-cone approximation. As evident, the nonlinearity of the tight-binding bands becomes important at high energies.

As graphene is heavily $n$-doped in our experiments, only the conduction band is relevant and we suppress the band index for brevity in the following.

\subsection{Carrier scattering}

In the Born approximation, the scattering rate due to random impurities is given by
\begin{equation}
    \frac{1}{\tau_{\bk}} = \frac{2\pi}{\hbar} N_\alpha \sum_{\bk'}
        \abs{V_{\bk\bk'}^\alpha}^2 \delta(\varepsilon_{\bk} - \varepsilon_{\bk'}),
\end{equation}
where $V_{\bk\bk'}^\alpha$ is the matrix elements of the individual impurity potentials, $V_\alpha(\br)=V_\alpha(\br-\mathbf{R}_\alpha)$, and $\mathbf{R}_\alpha$ is the position of the impurity in the primitive cell. Intra- and intervalley contributions to the scattering rate are separated out by confining the sum over $\bk'$ to, respectively, the same or the opposite valley of $\bk$.

\subsubsection{Charged Li adatoms}

The the charged Li adatoms are modelled by a point-charge impurities. To calculate the matrix element of the associated Coulomb scattering potential, we use the Bloch functions in Eq.~\eqref{eq:wf} in the definition of the matrix element,
\begin{align}
    \label{eq:Vmatrix}
    V_{\bk\bk'}^\text{Li} & = \int d\br\, \Psi_{\bk}^*(\br) V_\text{Li}(\br) \Psi_{\bk'}(\br) 
    = \sum_{\nu\nu'} \chi_{\nu \bk}^*\chi_{\nu' \bk'} 
           \int \! d\br\, \phi_{\nu \bk}^*(\br) V_\text{Li}(\br) \phi_{\nu' \bk'}(\br) .
\end{align}
To facilitate a semi-analytic evaluation of the matrix element, we express the Li impurity potential $V_\text{Li}(\br)$ by its 2D Fourier transform, i.e.
\begin{equation}
    V_\text{Li}(\br) = \frac{1}{A} \sum_{\bq,\mathbf{G}} e^{i(\bq+\mathbf{G})\cdot (\br-\mathbf{R}_\text{Li})} V_C(\bq+\mathbf{G}, z) ,
\end{equation}
where $A=N A_\text{cell}$ is the sample area, $\bq\in 1.$~BZ, $\mathbf{G}$ is a reciprocal lattice vector, and $\mathbf{R}_\text{Li}$ denotes the hollow position of the Li adatoms. The Fourier transform of the point-charge Coulomb potential is given by
\begin{equation}
    \label{eq:V_C}
    V_C(\bq, z) = \frac{e^2 Z}{2\epsilon_0 \kappa \varepsilon(\bq) q} e^{-q \abs{z-d}} ,
\end{equation}
where $d$ is the distance between the Li adatoms and the graphene layer, $Z$ is the valence of the Li adatoms, $\kappa=(\varepsilon_\text{SiC}+1)/2$ accounts for background screening by the substrate, and $\varepsilon(\bq)=1 - v(\abs{\bq})\chi(\bq)$ is the static dielectric function of doped graphene with $v(\abs{\bq})=\frac{e^2}{2\epsilon_0 \abs{\bq}}$ denoting the Fourier transform of the bare Coulomb potential in 2D, and $\chi(\bq)$ is the static polarizability here described by its analytic Dirac-cone form~\cite{s.Sarma:Carrier}.

Inserting in Eq.~\eqref{eq:Vmatrix}, we can approximate as
\begin{equation}
    \label{eq:V_Li}
    V_{\bk\bk'}^\text{Li} \approx \frac{1}{A} V_C(\bq, z=0) n_{\bk\bk'}^\text{Li}
    , \quad \bq = \bk'-\bk , \quad \mathbf{G} = \mathbf{0} ,
\end{equation}
where we have neglected umklapp processes involving $\mathbf{G}\neq \mathbf{0}$ Fourier components of the impurity potential, and the matrix element $n_{\bk\bk'}^\text{Li}$ is given by~\cite{Schliemann:Interacting}
\begin{align}
    n_{\bk\bk'}^\text{Li} & = \bra{\bk} e^{-i\bq\cdot (\br-\mathbf{R}_\text{Li})} \ket{\bk'} 
    = \sum_{\nu\nu'} \chi_{\nu \bk}^*\chi_{\nu' \bk'} \, \bra{\nu\bk} e^{-i\bq\cdot (\br-\mathbf{R}_\text{Li})} \ket{\nu'\bk'}
    \\ &
        \label{eq:form}
    \approx \delta_{\bk',\bk+\bq} 
        f(\bq) \times \sum_{\nu} \chi_{\nu \bk}^*\chi_{\nu \bk'} \, 
        e^{-i \bq\cdot(\mathbf{R}_\nu-\mathbf{R}_\text{Li})} .
\end{align}
Here, $f$ is a form factor given by the matrix element of the phase factor $\exp{(-i \bq \cdot\br)}$ with respect to a $p_z$ orbital at the origin,
\begin{equation}
  \label{eq:f}
  f(\bq) =  \int \! d\br\, e^{-i \bq \cdot\br} \abs{\phi_\nu(\br)}^2 .
\end{equation}
The integral can be evaluated analytically, and is~\cite{Schliemann:Interacting}
\begin{equation}
  \label{eq:f1}
  f(\bq) = \frac{1}{(1 + \abs{\bq}^2 a_0^2)^3} ,
\end{equation}
with $a_0 \approx a/(2\sqrt{10})$ and $a=2.46$~{\AA} is the lattice constant.

\subsubsection{Residual short-range disorder}

The residual disorder is modelled by the standard short-range impurity potential $V_\text{res}(\br)=V_0 \delta(\br - \mathbf{R}_i)$ where $V_0$ is the disorder strength. In a tight-binding description, this corresponds to a local shift of the onsite energy at the position of the impurity, and the matrix element simply becomes
\begin{equation}
    V_{\bk\bk'}^\textbf{res} = \frac{1}{A} V_0 .
\end{equation}

\subsection{Boltzmann transport calculations}

The following section outlines our calculations of the conductivity/mobility based on the linearized Boltzmann equation. 

The current density in the direction of the applied electric field $\mathbf{E}=E \hat{\mathbf{E}}$ is given by the out-of-equilibrium distribution function $f_{\bk}$ as
\begin{equation}
    \label{eq:j}
    j = \frac{g_s}{A} \sum_{\bk} (\mathbf{v}_{\bk}\cdot \hat{\mathbf{E}}) \delta f_{\bk} ,
\end{equation}
where $g_s=2$ is the spin degeneracy, $\mathbf{v}_{\bk}$ is the band velocity, and $\delta f_{\bk} = f_{\bk} - n_F(\varepsilon_{\bk})$ is the deviation away from the equilibrium Fermi-Dirac distribution $n_F$ to linear order in the applied field, $\delta f_{\bk}\propto E$. The longitudinal conductivity $\sigma=j/E$ follows then directly from Eq.~\eqref{eq:j}.

In the presence of elastic scattering, the linearized Boltzmann equation takes the form
\begin{equation}
    \label{eq:BE}
    q \mathbf{v}_{\bk} \cdot \mathbf{E} 
    \frac{\partial n_F}{\partial \varepsilon}\bigg\vert_{\varepsilon=\varepsilon_{\bk}}
    = -\sum_{\bk'} P_{\bk,\bk'} 
        \left[ \delta f_{\bk} - \delta f_{\bk'} \right]
\end{equation}
where the transition rate for elastic scattering off impurities in the Born approximation is given by
\begin{equation}
    P_{\bk,\bk'} = \frac{2\pi}{\hbar} \sum_\alpha N_\alpha \abs{V_{\bk\bk'}^\alpha}^2 
        \delta(\varepsilon_{\bk} - \varepsilon_{\bk'}) ,
\end{equation}
$N_\alpha$ is the number of impurities of type $\alpha$ with matrix element $V_{\bk\bk'}^\alpha$.

Here, we pursue a general numerical solution to the linearized Boltzmann equation which takes into account the anisotropi and nonlinearity of the tight-binding band structure. The Boltzmann equation~\eqref{eq:BE} can be recast as a matrix equation in the $\bk$ index,
\begin{equation}
    \label{eq:BE_matrix}
    \mathbf{C}\, \mathbf{\tilde{f}} = \mathbf{b}, \quad  
    \tilde{f}_{\bk} = \frac{\delta f_{\bk}}
                    {q \abs{\mathbf{E}} \frac{\partial n_F}{\partial \varepsilon}\big\vert_{\varepsilon=\varepsilon_{\bk}}}    ,
\end{equation}
which is solved for the vector $\mathbf{\tilde{f}}$, and where the matrix elements of the collision matrix $\mathbf{C}$ and the right-hand side are given by
\begin{equation}
    C_{\bk,\bk'} = - \delta_{\bk,\bk'} \sum_{\bk''} P_{\bk,\bk''} + P_{\bk,\bk'}
    \quad \text{and} \quad  
    b_{\bk} = \mathbf{v}_{\bk} \cdot \mathbf{\hat{E}} ,
\end{equation}
and $\mathbf{\hat{E}}=\mathbf{E}/\abs{\mathbf{E}}$ is a unit vector in the direction of the applied electric field.

We use a least-square method to solve the matrix equation~\eqref{eq:BE_matrix} appended with the additional particle-conserving constraint $\sum_{\bk} \delta f_{\bk}=0$ on the distribution function. The solution is based on a singular-value decomposition of the collision matrix, in which small singular values are set to zero to eliminate undesired contributions to the solution from a potentially finite-dimensional null space of the collision matrix.


\subsection{Dirac model}

The linearization of the tight-binding Hamiltonian in $\bk$ around the high-symmetry $K,K'$ points results in the well-known Dirac model approximation with linear dispersion $\varepsilon_\bk=\hbar v_F k$ and the valley-dependent eigenspinor $\chi_{\zeta\mathbf{k}}=\tfrac{1}{\sqrt{2}}(1,\zeta e^{i\zeta \theta_\mathbf{k}})^T$ with $\zeta=\pm 1$ denoting the valley index.

In the Dirac model, the calculation of the scattering rates and conductivity/mobility simplifies tremendously and can be done analytically. 

Our starting point is Matthiessen's rule for the mobility which applies at low temperatures ($T \ll T_F$), and which stats that the total inverse mobility can be obtained as
\begin{equation}
\label{eq:mob1}
	\mu_\text{tot}^{-1} = \mu_\text{res}^{-1} + \mu_\text{Li}^{-1}
\end{equation}
where $\mu_\alpha = \tfrac{\sigma_\alpha}{en}$, $\sigma_\alpha = \tfrac{e^2v_F^2}{2}D(E_F)\tau_\alpha(E_F)$ is the conductivity, $D(E_F)=\tfrac{2}{\hbar v_{F}}\sqrt{\tfrac{n}{\pi}}$ is graphene's density of states, and $\tau_\alpha$ are the relaxation times for the different scattering mechanisms. As a result, the relation between the total inverse mobility and the relaxation times can be written as
\begin{equation}
\label{eq:mob2}
	\frac{1}{\mu_\text{tot}} = \frac{\hbar \sqrt{\pi n}}{ev_{F}}\sum_{\alpha,v} \frac{1}{\tau_{\alpha,v}(E_F)}
\end{equation}
where $\alpha=\{\text{Li},\,\text{res}\}$ denotes the impurity type and $v=\{\ast,\,\text{i}\}$ represents intra- or intervalley scattering. The inverse relaxation time $\tau_{\alpha,v}$ is given by the expression in Eq.~(2) of the main manuscript with the replacment $1\rightarrow (1-\cos\theta_{\bk\bk'})$ in the $\bk$ integral.

The matrix element for intra- ($\zeta=\zeta'$) and intervalley ($\zeta\neq\zeta'$) scattering is here expressed as
\begin{equation}
	V_{\mathbf{k}\mathbf{k}'}^\alpha(\zeta,\zeta') = \bra{\chi_{\zeta\mathbf{k}}}\hat{V}_\alpha(\bq)\ket{\chi_{\zeta'\mathbf{k}'}},
\end{equation}
where $\hat{V}_\alpha(\bq)$ is the scattering potential and $\bq=\mathbf{k}-\mathbf{k}'$.

\subsubsection{Residual short-range disorder}

For random residual disorder distributed equally on the A and B sublattice, $\hat{V}_\text{res} \propto \sigma_0 \pm \sigma_z$, where $\sigma_0$ is the $2\times 2$ identity matrix and $\sigma_z$ is the Pauli matrix. The absolute square of the matrix element then becomes
\begin{align}
	\abs{V_{\bk\bk'}^\text{res}(\zeta,\zeta')}^2 
    & = V_0^2 \abs{\bra{\chi_{\zeta\bk}} \sigma_0 \pm \sigma_z \ket{\chi_{\zeta'\bk'}}}^2
     = V_0^2  ,
\end{align}
with different disorder strengths for intra- and intervalley scattering, respectively. When the matrix element is independent on $\bk,\bk'$, the evaluation of the relaxation time is trivial,
\begin{align}
  \label{eq:tau_short}
  \frac{1}{\tau_\text{res}(E_F)} & = \frac{n_\text{res} V_{0}^2}{2\pi\hbar} \frac{k_F}{\hbar v_F}
      \int_0^{2\pi} \! d\theta_{\bk\bk'} \, 
      \big[ 1 - \cos\theta_{\bk\bk'} \big] \nonumber \\
  & = \frac{n_\text{res} V_{0}^2}{\hbar} \frac{E_F}{(\hbar v_F)^2}  .
\end{align}

\subsubsection{Scattering by Li adatoms}

As shown by DFT calculations in Ref.~\onlinecite{s.Kristen}, the scattering potential of the Li adatoms is dominated by the long-range Coulomb potential arising due to their net positive charge.
Because of the hollow site position of the Li adatoms, their impurity potential does not break the sublattice symmetry and the scattering potential hence becomes diagonal in the sublattice basis, i.e.
$\hat{V}_\text{Li}(\bq) \propto \sigma_0$.

The absolute square of the matrix element for Li-induced intravalley scattering becomes
\begin{align}
\label{eq:VC_intra}	
	\abs{V_{\bk\bk'}^\text{Li}(\zeta,\zeta)}^2 
    & = V_C(q)^2 \abs{\bra{\chi_{\zeta\bk}}\sigma_0 \ket{\chi_{\zeta\bk'}}}^2
    \nonumber \\
    & = V_C(q)^2 \cos^2(\theta_{\bk\bk'}/2)
\end{align}
where $V_C$ is the Fourier transform of the screened Coulomb potential in Eq.~\eqref{eq:V_C} above.

Inserting in the expression for the relaxation time in the main text [Eq.~2], we find the following formula for the Li-induced intravalley scattering rate
\begin{align}
  \label{eq:tau_C}
  \frac{1}{\tau_{\text{Li}}(E_F)} & = \frac{n_\text{Li}}{2\pi\hbar} \frac{k_F}{\hbar v_F}
      \left( \frac{Z_\text{Li}e^2}{2\epsilon_0\kappa} \right)^2 
      \nonumber \\
      & \quad \times \int_0^{2\pi} \! d\theta_{\bk\bk'} \, e^{-2q d}
      \frac{\cos^2{\theta_{\bk\bk'}/2}}{(q + q_\text{TF})^2}
                    \big[ 1 - \cos\theta_{\bk\bk'} \big] \nonumber \\
                  & = \frac{n_\text{Li} }{2\pi\hbar}
                      \left( \frac{Z_\text{Li}e^2}{2\epsilon_0\kappa} \right)^2
                      \frac{2}{E_F}  \int_0^1 \! dx \,e^{-4k_F x d}
                    \frac{x^2 \sqrt{1- x^2} }{(x + q_0)^2} 
                    \nonumber \\
\end{align}
where $q = 2k_F\sin\theta_{\bk\bk'}/2$, $x=\sin\theta_{\bk\bk'}/2$, $q_0 = q_\text{TF} / 2k_F$, and $q_\text{TF}=4e^2k_F/(4\pi\epsilon_0\kappa\hbar v_F)$ is the Thomas-Fermi wave vector.

When $d=0$, the exact solution of the integral is given by 
\begin{align}
\label{eq:tau_C,d=0}
 \frac{1}{\tau_{\text{Li}}(E_F)} & =\frac{n_\text{Li} }{\pi\hbar}
                      \left( \frac{Z_\text{Li} e^2}{2\epsilon_0 \kappa} \right)^2
                      F(q_0) \frac{1}{E_F} \\
	F(q_0) & = \frac{\pi}{4} +  3 q_0 - \frac{3\pi}{2} q_0^2 + 
             \frac{q_0\left[3 q_0^2 - 2 \right] \mathrm{arccos}
             	\left(\tfrac{1}{q_0}\right)}{\sqrt{q_0^2-1}} ,
\end{align}
in agreement with Ref.~\onlinecite{sAdamPNAS2007}.

For the intervalley matrix element ($\zeta\neq\zeta'$) we have
\begin{align}
	\label{eq:VC_inter}
	\abs{V_{\bk\bk'}^\text{Li}(\zeta,\zeta')}^2 
    & = V_C(q)^2 \abs{\bra{\chi_{\zeta\bk}}\sigma_0 \ket{\chi_{\zeta'\bk'}}}^2
    \nonumber \\
    & = V_C(q)^2 \sin^2(\theta_{\bk\bk'}/2) ,
\end{align}
which in contrast to intravalley scattering suppresses forward scattering instead of backscattering. 
Writing the intervalley scattering wave vector as $\bq' = \bK + \bq$, the 2D dielectric function becomes
\begin{align}
	\varepsilon(q') & = 1 + \frac{q_\text{TF}}{\abs{\bK + \bq}}
    \approx 1 + \frac{q_\text{TF}}{\abs{\bK}} \sim 1 ,
\end{align}
which implies that the intervalley components of the Coulomb potential are mainly screened by the dielectric environment ($\kappa$). The screened Coulomb potential for intervalley scattering thus becomes
\begin{align}
	\label{eq:VCK}
	V_C(q') & \approx \frac{Z_\text{Li} e^2}{2\epsilon_0 \kappa} \frac{e^{-q'd}}{\abs{\bq'}} ,
\end{align}
which to a good approximation can be assumed constant, $V_{C,0}\equiv V_C(q'=\abs{\bK})$. The factor $e^{-\abs{\bK}d}$ in the intervalley Coulomb potential makes the Li-induced intervalley rate two orders of magnitude smaller than the one for residual intervalley scattering. Therefore, the Li-induced intervalley rate can be ignored.


\subsection{Fitting procedure and parameters}

\begin{table}[!h]
\begin{tabular}{lcc}
\hline
Parameter &  Symbol  & Value  \\
\hline
Fermi velocity                 &   $v_F$                    &   $10^6\,\mathrm{m}/\mathrm{s}$             \\
Li valence                     &   $Z_\text{Li}$            &   +0.9             \\
Li-graphene distance               &   $d$                      &   1.78~\AA             \\
Substrate screening            &   $\varepsilon_\text{SiC}$ &   13.5             \\
\hline
Residual short-range disorder  &                            &                     \\
\hline
Density                        &   $n_\text{res}$           &   $10^{12}$~$\mathrm{cm}^{-2}$ \\
SiC4-700K                      &                            &                               \\
Intravalley                    &   $V_{0,\ast}$         &   $75\,\mathrm{eV}\,\text{\AA}^2$          \\
Intervalley                    &   $V_{0,\text{i}}$         &   $29\,\mathrm{eV}\,\text{\AA}^2$           \\
SiC3-900K                      &                            &                                              \\
Intravalley                    &   $V_{0,\ast}$         &   $107\,\mathrm{eV}\,\text{\AA}^2$          \\
Intervalley                    &   $V_{0,\text{i}}$         &   $21\,\mathrm{eV}\,\text{\AA}^2$           \\
\end{tabular}
\caption{Material and disorder parameters used in the calculation of the scattering rates and conductivity.}
\label{tab:parameters}
\end{table}
This section describes the fitting procedure applied to obtain the parameters for the residual disorder (i.e., density and disorder strengths) used in the calculation of the theoretical lines in Fig.~3. 

\begin{itemize}
\item First, we fix the intervalley disorder strength by fitting the intervalley scattering rate at $\Delta n=0$, assuming a density of residual disorder of $n_\text{res}=10^{12}~\mathrm{cm}^{-2}$.
\item Secondly, we fix the intravalley disorder strength by fitting to the conductivity/mobility at $\Delta n=0$. 
\item All parameters enterning the matrix element of the Li scattering potential in Eq.~\eqref{eq:V_Li} have been inferred from DFT calculations~\cite{s.Kristen}. Except for the dielectric constant of SiC substrate for which we used $\varepsilon_\text{SiC}=13.5$~\cite{s.SiC.dielectric.constant.Sci.Rep.2012,s.Bart.Tl.Graphene.long.and.short.range.scatt.Nano.Lett.2015}. 
\end{itemize}
The parameters used for the two devices in Fig.~3(b) of the main manuscript are summarized in Table~\ref{tab:parameters}.

\subsection{Discussion about lack of superconductivity in Li-doped graphene}
\label{sec:NO SC Discussion}

How can we understand the lack of a conductivity upturn as low as 3 or 4 K for Li-doped graphene, while Ref. \cite{s.LiSC.NatPhys2012} predicted a T$_{C}$=8.1 K and Ref. \cite{s.Bart} reported evidence of a temperature-dependent pairing gap corresponding to a $T_{C}\simeq$5.9 K? We consider three possibilities:  
\begin{enumerate}
\item The transition to superconductivity in quasi-2D films is governed by superconducting fluctuations \cite{s.SC.Ca.bilayer.graphene.ACSNano.2016, s.2D.superconducting.fluctuations.PhysRevLett.1971, s.low.D.SC.Phys.Lett.A.1968}, and is not as abrupt as it is for 3D materials.  It is in principle possible that a gradual reduction in resistance with decreasing T could be hidden on top of the increasing resistance due to weak localization and electron-electron interactions.  In that case, however, one would expect significantly modified magnetoresistance curves, reflecting weak localization on top of magnetic field suppression of incipient superconductivity. This was not observed.

\item Thermal fluctuations can suppress superconductivity in 2D systems via the Berezinskii-Kosterlitz-Thouless (BKT) transition. In this case, a system may possess a pseudogap without showing any suppression of resistance \cite{s.BKT.Nat.Phys.2007, s.BKT.graphene.SC.PhysRevB.2009, s.Richter.Nature2013}. 
The BKT scenario has been observed experimentally for proximity-induced superconductivity on graphene \cite{s.BKT.graphene.proximity.SC.PhysRevLett.2010, s.BKT.graphene.proximity.SC.Nat.Phys.2014}, and predicted theoretically for superconductivity in doped graphene \cite{s.BKT.graphene.SC.PhysRevB.2009}.
To estimate the importance of this effect, we use an expression for the BKT transition temperature that is well established in metals: $k_BT_{BKT}=d\Phi _0^2\rho_c/8\pi\mu _0$ where $d$ is the graphene thickness, $\Phi _0$=h/2e is the flux quantum, and $\rho_c$ is the superfluid density\cite{s.BKT.Nat.Phys.2007, s.BKT.Beasley.PhysRevLett.1979}. Using Homes' law \cite{s.Homes.law.Nat.2004} to estimate superfluid density in the SiC sample after Li deposition, $\rho_c\sim 120 \sigma_N T_c/d$ where $\sigma_N=0.007$ is the normal state 2D conductivity in $\Omega ^{-1}$, $d=3.4\times 10^{-8}$ is the graphene's ``thickness" in cm, and $T_c=5.9$ K is the critical temperature, we find $\rho_c\sim 1.5\times 10^{8}$ cm$^{-2}$ gives $T_{BKT}\sim 5$ K. Given the significant approximations involved in the above calculation, the fact that $T_c$ and $T_{BKT}$ are so similar  shows that a BKT-induced suppression of $T_c$ must be considered, calling for further measurements at significantly lower temperatures.

\item ARPES-detected signatures of superconductivity due to Li adatoms were observed only for some SiC samples, and only after repeated annealing operations monitored by the sharpness of the graphene band structure \cite{s.privcom}. It is possible that superconductivity by Li adatoms requires a specific graphene condition that was not realized in our experiments. The SiC data reported here are not for the specific chip used in Ref.~\onlinecite{s.Bart}. We first measured that chip but found an extremely anisotropic resistance; the SiC1 sample reported here was grown later in the same chamber, aiming for more optimal growth parameters. Unfortunately, due to the low resistance of SiC1 substrate at room temperature, it was not possible to anneal its graphene in our heater stage. For performing high-temperature annealing, we used SiC2,3,4 that were cut from a commercially available epitaxial monolayer graphene. These samples may have been grown under different conditions compared to SiC1.

\end{enumerate}


\begin{thebibliography}{44}%
\makeatletter
\providecommand \@ifxundefined [1]{%
 \@ifx{#1\undefined}
}%
\providecommand \@ifnum [1]{%
 \ifnum #1\expandafter \@firstoftwo
 \else \expandafter \@secondoftwo
 \fi
}%
\providecommand \@ifx [1]{%
 \ifx #1\expandafter \@firstoftwo
 \else \expandafter \@secondoftwo
 \fi
}%
\providecommand \natexlab [1]{#1}%
\providecommand \enquote  [1]{``#1''}%
\providecommand \bibnamefont  [1]{#1}%
\providecommand \bibfnamefont [1]{#1}%
\providecommand \citenamefont [1]{#1}%
\providecommand \href@noop [0]{\@secondoftwo}%
\providecommand \href [0]{\begingroup \@sanitize@url \@href}%
\providecommand \@href[1]{\@@startlink{#1}\@@href}%
\providecommand \@@href[1]{\endgroup#1\@@endlink}%
\providecommand \@sanitize@url [0]{\catcode `\\12\catcode `\$12\catcode
  `\&12\catcode `\#12\catcode `\^12\catcode `\_12\catcode `\%12\relax}%
\providecommand \@@startlink[1]{}%
\providecommand \@@endlink[0]{}%
\providecommand \url  [0]{\begingroup\@sanitize@url \@url }%
\providecommand \@url [1]{\endgroup\@href {#1}{\urlprefix }}%
\providecommand \urlprefix  [0]{URL }%
\providecommand \Eprint [0]{\href }%
\providecommand \doibase [0]{http://dx.doi.org/}%
\providecommand \selectlanguage [0]{\@gobble}%
\providecommand \bibinfo  [0]{\@secondoftwo}%
\providecommand \bibfield  [0]{\@secondoftwo}%
\providecommand \translation [1]{[#1]}%
\providecommand \BibitemOpen [0]{}%
\providecommand \bibitemStop [0]{}%
\providecommand \bibitemNoStop [0]{.\EOS\space}%
\providecommand \EOS [0]{\spacefactor3000\relax}%
\providecommand \BibitemShut  [1]{\csname bibitem#1\endcsname}%
\let\auto@bib@innerbib\@empty
\bibitem [{\citenamefont {Profeta}\ \emph {et~al.}(2012)\citenamefont
  {Profeta}, \citenamefont {Calandra},\ and\ \citenamefont
  {Mauri}}]{LiSC.NatPhys2012}%
  \BibitemOpen
  \bibfield  {author} {\bibinfo {author} {\bibfnamefont {G.}~\bibnamefont
  {Profeta}}, \bibinfo {author} {\bibfnamefont {M.}~\bibnamefont {Calandra}}, \
  and\ \bibinfo {author} {\bibfnamefont {F.}~\bibnamefont {Mauri}},\ }\href
  {\doibase 10.1038/nphys2181} {\bibfield  {journal} {\bibinfo  {journal} {Nat
  Phys}\ }\textbf {\bibinfo {volume} {8}},\ \bibinfo {pages} {131} (\bibinfo
  {year} {2012})}\BibitemShut {NoStop}%
\bibitem [{\citenamefont {Flores-Livas}\ and\ \citenamefont
  {Sanna}(2015)}]{SC.Alkali.honeycomb.PhysRevB.2015}%
  \BibitemOpen
  \bibfield  {author} {\bibinfo {author} {\bibfnamefont {J.~A.}\ \bibnamefont
  {Flores-Livas}}\ and\ \bibinfo {author} {\bibfnamefont {A.}~\bibnamefont
  {Sanna}},\ }\href {\doibase 10.1103/PhysRevB.91.054508} {\bibfield  {journal}
  {\bibinfo  {journal} {Phys. Rev. B}\ }\textbf {\bibinfo {volume} {91}},\
  \bibinfo {pages} {054508} (\bibinfo {year} {2015})}\BibitemShut {NoStop}%
\bibitem [{\citenamefont {Uchoa}\ and\ \citenamefont
  {Castro~Neto}(2007)}]{Plasmon.SC.PhysRevLett.2007}%
  \BibitemOpen
  \bibfield  {author} {\bibinfo {author} {\bibfnamefont {B.}~\bibnamefont
  {Uchoa}}\ and\ \bibinfo {author} {\bibfnamefont {A.~H.}\ \bibnamefont
  {Castro~Neto}},\ }\href {\doibase 10.1103/PhysRevLett.98.146801} {\bibfield
  {journal} {\bibinfo  {journal} {Phys. Rev. Lett.}\ }\textbf {\bibinfo
  {volume} {98}},\ \bibinfo {pages} {146801} (\bibinfo {year}
  {2007})}\BibitemShut {NoStop}%
\bibitem [{\citenamefont {Nandkishore}\ \emph {et~al.}(2012)\citenamefont
  {Nandkishore}, \citenamefont {Levitov},\ and\ \citenamefont
  {Chubukov}}]{Chiral.SC.Nat.Phys2012}%
  \BibitemOpen
  \bibfield  {author} {\bibinfo {author} {\bibfnamefont {R.}~\bibnamefont
  {Nandkishore}}, \bibinfo {author} {\bibfnamefont {L.~S.}\ \bibnamefont
  {Levitov}}, \ and\ \bibinfo {author} {\bibfnamefont {A.~V.}\ \bibnamefont
  {Chubukov}},\ }\href {\doibase 10.1038/nphys2208} {\bibfield  {journal}
  {\bibinfo  {journal} {Nat Phys}\ }\textbf {\bibinfo {volume} {8}},\ \bibinfo
  {pages} {158} (\bibinfo {year} {2012})}\BibitemShut {NoStop}%
\bibitem [{\citenamefont {Hong}\ \emph {et~al.}(2012)\citenamefont {Hong},
  \citenamefont {Zou}, \citenamefont {Wang}, \citenamefont {Cheng},\ and\
  \citenamefont {Zhu}}]{Flourine.PhysRevLett2012}%
  \BibitemOpen
  \bibfield  {author} {\bibinfo {author} {\bibfnamefont {X.}~\bibnamefont
  {Hong}}, \bibinfo {author} {\bibfnamefont {K.}~\bibnamefont {Zou}}, \bibinfo
  {author} {\bibfnamefont {B.}~\bibnamefont {Wang}}, \bibinfo {author}
  {\bibfnamefont {S.-H.}\ \bibnamefont {Cheng}}, \ and\ \bibinfo {author}
  {\bibfnamefont {J.}~\bibnamefont {Zhu}},\ }\href {\doibase
  10.1103/PhysRevLett.108.226602} {\bibfield  {journal} {\bibinfo  {journal}
  {Phys. Rev. Lett.}\ }\textbf {\bibinfo {volume} {108}},\ \bibinfo {pages}
  {226602} (\bibinfo {year} {2012})}\BibitemShut {NoStop}%
\bibitem [{\citenamefont {Eelbo}\ \emph {et~al.}(2013)\citenamefont {Eelbo},
  \citenamefont {Wa\ifmmode~\acute{s}\else \'{s}\fi{}niowska}, \citenamefont
  {Thakur}, \citenamefont {Gyamfi}, \citenamefont {Sachs}, \citenamefont
  {Wehling}, \citenamefont {Forti}, \citenamefont {Starke}, \citenamefont
  {Tieg}, \citenamefont {Lichtenstein},\ and\ \citenamefont
  {Wiesendanger}}]{3dTransition.PhysRevLett.2013}%
  \BibitemOpen
  \bibfield  {author} {\bibinfo {author} {\bibfnamefont {T.}~\bibnamefont
  {Eelbo}}, \bibinfo {author} {\bibfnamefont {M.}~\bibnamefont
  {Wa\ifmmode~\acute{s}\else \'{s}\fi{}niowska}}, \bibinfo {author}
  {\bibfnamefont {P.}~\bibnamefont {Thakur}}, \bibinfo {author} {\bibfnamefont
  {M.}~\bibnamefont {Gyamfi}}, \bibinfo {author} {\bibfnamefont
  {B.}~\bibnamefont {Sachs}}, \bibinfo {author} {\bibfnamefont {T.~O.}\
  \bibnamefont {Wehling}}, \bibinfo {author} {\bibfnamefont {S.}~\bibnamefont
  {Forti}}, \bibinfo {author} {\bibfnamefont {U.}~\bibnamefont {Starke}},
  \bibinfo {author} {\bibfnamefont {C.}~\bibnamefont {Tieg}}, \bibinfo {author}
  {\bibfnamefont {A.~I.}\ \bibnamefont {Lichtenstein}}, \ and\ \bibinfo
  {author} {\bibfnamefont {R.}~\bibnamefont {Wiesendanger}},\ }\href {\doibase
  10.1103/PhysRevLett.110.136804} {\bibfield  {journal} {\bibinfo  {journal}
  {Phys. Rev. Lett.}\ }\textbf {\bibinfo {volume} {110}},\ \bibinfo {pages}
  {136804} (\bibinfo {year} {2013})}\BibitemShut {NoStop}%
\bibitem [{\citenamefont {Weeks}\ \emph {et~al.}(2011)\citenamefont {Weeks},
  \citenamefont {Hu}, \citenamefont {Alicea}, \citenamefont {Franz},\ and\
  \citenamefont {Wu}}]{Franz.PhysRevX.1}%
  \BibitemOpen
  \bibfield  {author} {\bibinfo {author} {\bibfnamefont {C.}~\bibnamefont
  {Weeks}}, \bibinfo {author} {\bibfnamefont {J.}~\bibnamefont {Hu}}, \bibinfo
  {author} {\bibfnamefont {J.}~\bibnamefont {Alicea}}, \bibinfo {author}
  {\bibfnamefont {M.}~\bibnamefont {Franz}}, \ and\ \bibinfo {author}
  {\bibfnamefont {R.}~\bibnamefont {Wu}},\ }\href {\doibase
  10.1103/PhysRevX.1.021001} {\bibfield  {journal} {\bibinfo  {journal} {Phys.
  Rev. X}\ }\textbf {\bibinfo {volume} {1}},\ \bibinfo {pages} {021001}
  (\bibinfo {year} {2011})}\BibitemShut {NoStop}%
\bibitem [{\citenamefont {Hu}\ \emph {et~al.}(2012)\citenamefont {Hu},
  \citenamefont {Alicea}, \citenamefont {Wu},\ and\ \citenamefont
  {Franz}}]{Franz.PhysRevLett.2012}%
  \BibitemOpen
  \bibfield  {author} {\bibinfo {author} {\bibfnamefont {J.}~\bibnamefont
  {Hu}}, \bibinfo {author} {\bibfnamefont {J.}~\bibnamefont {Alicea}}, \bibinfo
  {author} {\bibfnamefont {R.}~\bibnamefont {Wu}}, \ and\ \bibinfo {author}
  {\bibfnamefont {M.}~\bibnamefont {Franz}},\ }\href {\doibase
  10.1103/PhysRevLett.109.266801} {\bibfield  {journal} {\bibinfo  {journal}
  {Phys. Rev. Lett.}\ }\textbf {\bibinfo {volume} {109}},\ \bibinfo {pages}
  {266801} (\bibinfo {year} {2012})}\BibitemShut {NoStop}%
\bibitem [{\citenamefont {Chen}\ \emph {et~al.}(2008)\citenamefont {Chen},
  \citenamefont {Jang}, \citenamefont {Adam}, \citenamefont {Fuhrer},
  \citenamefont {Williams},\ and\ \citenamefont {Ishigami}}]{K.Nat.Phys.2008}%
  \BibitemOpen
  \bibfield  {author} {\bibinfo {author} {\bibfnamefont {J.-H.}\ \bibnamefont
  {Chen}}, \bibinfo {author} {\bibfnamefont {C.}~\bibnamefont {Jang}}, \bibinfo
  {author} {\bibfnamefont {S.}~\bibnamefont {Adam}}, \bibinfo {author}
  {\bibfnamefont {M.~S.}\ \bibnamefont {Fuhrer}}, \bibinfo {author}
  {\bibfnamefont {E.~D.}\ \bibnamefont {Williams}}, \ and\ \bibinfo {author}
  {\bibfnamefont {M.}~\bibnamefont {Ishigami}},\ }\href {\doibase
  10.1038/nphys935} {\bibfield  {journal} {\bibinfo  {journal} {Nat Phys}\
  }\textbf {\bibinfo {volume} {4}},\ \bibinfo {pages} {377} (\bibinfo {year}
  {2008})}\BibitemShut {NoStop}%
\bibitem [{\citenamefont {Chan}\ \emph {et~al.}(2008)\citenamefont {Chan},
  \citenamefont {Neaton},\ and\ \citenamefont
  {Cohen}}]{DFT.Metals.PhysRevB.2008}%
  \BibitemOpen
  \bibfield  {author} {\bibinfo {author} {\bibfnamefont {K.~T.}\ \bibnamefont
  {Chan}}, \bibinfo {author} {\bibfnamefont {J.~B.}\ \bibnamefont {Neaton}}, \
  and\ \bibinfo {author} {\bibfnamefont {M.~L.}\ \bibnamefont {Cohen}},\ }\href
  {\doibase 10.1103/PhysRevB.77.235430} {\bibfield  {journal} {\bibinfo
  {journal} {Phys. Rev. B}\ }\textbf {\bibinfo {volume} {77}},\ \bibinfo
  {pages} {235430} (\bibinfo {year} {2008})}\BibitemShut {NoStop}%
\bibitem [{\citenamefont {Liu}\ \emph {et~al.}(2011)\citenamefont {Liu},
  \citenamefont {Wang}, \citenamefont {Yao}, \citenamefont {Lu}, \citenamefont
  {Hupalo}, \citenamefont {Tringides},\ and\ \citenamefont
  {Ho}}]{Bonding.Metal.adatom.graphene.PhysRevB.2011}%
  \BibitemOpen
  \bibfield  {author} {\bibinfo {author} {\bibfnamefont {X.}~\bibnamefont
  {Liu}}, \bibinfo {author} {\bibfnamefont {C.~Z.}\ \bibnamefont {Wang}},
  \bibinfo {author} {\bibfnamefont {Y.~X.}\ \bibnamefont {Yao}}, \bibinfo
  {author} {\bibfnamefont {W.~C.}\ \bibnamefont {Lu}}, \bibinfo {author}
  {\bibfnamefont {M.}~\bibnamefont {Hupalo}}, \bibinfo {author} {\bibfnamefont
  {M.~C.}\ \bibnamefont {Tringides}}, \ and\ \bibinfo {author} {\bibfnamefont
  {K.~M.}\ \bibnamefont {Ho}},\ }\href {\doibase 10.1103/PhysRevB.83.235411}
  {\bibfield  {journal} {\bibinfo  {journal} {Phys. Rev. B}\ }\textbf {\bibinfo
  {volume} {83}},\ \bibinfo {pages} {235411} (\bibinfo {year}
  {2011})}\BibitemShut {NoStop}%
\bibitem [{\citenamefont {Katoch}\ and\ \citenamefont
  {Ishigami}(2012)}]{Ca.SolidStateCommunications2012}%
  \BibitemOpen
  \bibfield  {author} {\bibinfo {author} {\bibfnamefont {J.}~\bibnamefont
  {Katoch}}\ and\ \bibinfo {author} {\bibfnamefont {M.}~\bibnamefont
  {Ishigami}},\ }\href {\doibase http://dx.doi.org/10.1016/j.ssc.2011.11.003}
  {\bibfield  {journal} {\bibinfo  {journal} {Solid State Communications}\
  }\textbf {\bibinfo {volume} {152}},\ \bibinfo {pages} {60 } (\bibinfo {year}
  {2012})}\BibitemShut {NoStop}%
\bibitem [{\citenamefont {Yan}\ and\ \citenamefont
  {Fuhrer}(2011)}]{K.PhysRevLett.2011}%
  \BibitemOpen
  \bibfield  {author} {\bibinfo {author} {\bibfnamefont {J.}~\bibnamefont
  {Yan}}\ and\ \bibinfo {author} {\bibfnamefont {M.~S.}\ \bibnamefont
  {Fuhrer}},\ }\href {\doibase 10.1103/PhysRevLett.107.206601} {\bibfield
  {journal} {\bibinfo  {journal} {Phys. Rev. Lett.}\ }\textbf {\bibinfo
  {volume} {107}},\ \bibinfo {pages} {206601} (\bibinfo {year}
  {2011})}\BibitemShut {NoStop}%
\bibitem [{\citenamefont {Ludbrook}\ \emph {et~al.}(2015)\citenamefont
  {Ludbrook}, \citenamefont {Levy}, \citenamefont {Nigge}, \citenamefont
  {Zonno}, \citenamefont {Schneider}, \citenamefont {Dvorak}, \citenamefont
  {Veenstra}, \citenamefont {Zhdanovich}, \citenamefont {Wong}, \citenamefont
  {Dosanjh}, \citenamefont {Stra{\ss}er}, \citenamefont {St\"ohr},
  \citenamefont {Forti}, \citenamefont {Ast}, \citenamefont {Starke},\ and\
  \citenamefont {Damascelli}}]{Bart}%
  \BibitemOpen
  \bibfield  {author} {\bibinfo {author} {\bibfnamefont {B.~M.}\ \bibnamefont
  {Ludbrook}}, \bibinfo {author} {\bibfnamefont {G.}~\bibnamefont {Levy}},
  \bibinfo {author} {\bibfnamefont {P.}~\bibnamefont {Nigge}}, \bibinfo
  {author} {\bibfnamefont {M.}~\bibnamefont {Zonno}}, \bibinfo {author}
  {\bibfnamefont {M.}~\bibnamefont {Schneider}}, \bibinfo {author}
  {\bibfnamefont {D.~J.}\ \bibnamefont {Dvorak}}, \bibinfo {author}
  {\bibfnamefont {C.~N.}\ \bibnamefont {Veenstra}}, \bibinfo {author}
  {\bibfnamefont {S.}~\bibnamefont {Zhdanovich}}, \bibinfo {author}
  {\bibfnamefont {D.}~\bibnamefont {Wong}}, \bibinfo {author} {\bibfnamefont
  {P.}~\bibnamefont {Dosanjh}}, \bibinfo {author} {\bibfnamefont
  {C.}~\bibnamefont {Stra{\ss}er}}, \bibinfo {author} {\bibfnamefont
  {A.}~\bibnamefont {St\"ohr}}, \bibinfo {author} {\bibfnamefont
  {S.}~\bibnamefont {Forti}}, \bibinfo {author} {\bibfnamefont {C.~R.}\
  \bibnamefont {Ast}}, \bibinfo {author} {\bibfnamefont {U.}~\bibnamefont
  {Starke}}, \ and\ \bibinfo {author} {\bibfnamefont {A.}~\bibnamefont
  {Damascelli}},\ }\href {\doibase 10.1073/pnas.1510435112} {\bibfield
  {journal} {\bibinfo  {journal} {Proceedings of the National Academy of
  Sciences}\ }\textbf {\bibinfo {volume} {112}},\ \bibinfo {pages} {11795}
  (\bibinfo {year} {2015})}\BibitemShut {NoStop}%
\bibitem [{\citenamefont {McChesney}\ \emph {et~al.}(2010)\citenamefont
  {McChesney}, \citenamefont {Bostwick}, \citenamefont {Ohta}, \citenamefont
  {Seyller}, \citenamefont {Horn}, \citenamefont {Gonz{\'a}lez},\ and\
  \citenamefont {Rotenberg}}]{Rotenberg.Extended}%
  \BibitemOpen
  \bibfield  {author} {\bibinfo {author} {\bibfnamefont {J.~L.}\ \bibnamefont
  {McChesney}}, \bibinfo {author} {\bibfnamefont {A.}~\bibnamefont {Bostwick}},
  \bibinfo {author} {\bibfnamefont {T.}~\bibnamefont {Ohta}}, \bibinfo {author}
  {\bibfnamefont {T.}~\bibnamefont {Seyller}}, \bibinfo {author} {\bibfnamefont
  {K.}~\bibnamefont {Horn}}, \bibinfo {author} {\bibfnamefont {J.}~\bibnamefont
  {Gonz{\'a}lez}}, \ and\ \bibinfo {author} {\bibfnamefont {E.}~\bibnamefont
  {Rotenberg}},\ }\href@noop {} {\bibfield  {journal} {\bibinfo  {journal}
  {Phys. Rev. Lett.}\ }\textbf {\bibinfo {volume} {104}},\ \bibinfo {pages}
  {136803} (\bibinfo {year} {2010})}\BibitemShut {NoStop}%
\bibitem [{\citenamefont {Fedorov}\ \emph {et~al.}(2014)\citenamefont
  {Fedorov}, \citenamefont {Verbitskiy}, \citenamefont {Haberer}, \citenamefont
  {Struzzi}, \citenamefont {Petaccia}, \citenamefont {Usachov}, \citenamefont
  {Vilkov}, \citenamefont {Vyalikh}, \citenamefont {Fink}, \citenamefont
  {Knupfer}, \citenamefont {B{\"u}chner},\ and\ \citenamefont
  {Gr{\"u}neis}}]{Gruneis.Observation}%
  \BibitemOpen
  \bibfield  {author} {\bibinfo {author} {\bibfnamefont {A.~V.}\ \bibnamefont
  {Fedorov}}, \bibinfo {author} {\bibfnamefont {N.~I.}\ \bibnamefont
  {Verbitskiy}}, \bibinfo {author} {\bibfnamefont {D.}~\bibnamefont {Haberer}},
  \bibinfo {author} {\bibfnamefont {C.}~\bibnamefont {Struzzi}}, \bibinfo
  {author} {\bibfnamefont {L.}~\bibnamefont {Petaccia}}, \bibinfo {author}
  {\bibfnamefont {D.}~\bibnamefont {Usachov}}, \bibinfo {author} {\bibfnamefont
  {O.~Y.}\ \bibnamefont {Vilkov}}, \bibinfo {author} {\bibfnamefont {D.~V.}\
  \bibnamefont {Vyalikh}}, \bibinfo {author} {\bibfnamefont {J.}~\bibnamefont
  {Fink}}, \bibinfo {author} {\bibfnamefont {M.}~\bibnamefont {Knupfer}},
  \bibinfo {author} {\bibfnamefont {B.}~\bibnamefont {B{\"u}chner}}, \ and\
  \bibinfo {author} {\bibfnamefont {A.}~\bibnamefont {Gr{\"u}neis}},\
  }\href@noop {} {\bibfield  {journal} {\bibinfo  {journal} {Nat. Commun.}\
  }\textbf {\bibinfo {volume} {5}},\ \bibinfo {pages} {3257} (\bibinfo {year}
  {2014})}\BibitemShut {NoStop}%
\bibitem [{\citenamefont {Kaasbjerg}\ and\ \citenamefont
  {Jauho}(2019)}]{Kristen}%
  \BibitemOpen
  \bibfield  {author} {\bibinfo {author} {\bibfnamefont {K.}~\bibnamefont
  {Kaasbjerg}}\ and\ \bibinfo {author} {\bibfnamefont {A.-P.}\ \bibnamefont
  {Jauho}},\ }\href@noop {} {\bibfield  {journal} {\bibinfo  {journal} {Phys.
  Rev. XXX}\ }\textbf {\bibinfo {volume} {XXX}},\ \bibinfo {pages} {XXX}
  (\bibinfo {year} {2019})},\ \bibinfo {note} {arXiv:1904.08191}\BibitemShut
  {NoStop}%
\bibitem [{\citenamefont {K{\"u}hne}\ \emph {et~al.}(2017)\citenamefont
  {K{\"u}hne}, \citenamefont {Paolucci}, \citenamefont {Popovic}, \citenamefont
  {Ostrovsky}, \citenamefont {Maier},\ and\ \citenamefont {Smet}}]{NNano}%
  \BibitemOpen
  \bibfield  {author} {\bibinfo {author} {\bibfnamefont {M.}~\bibnamefont
  {K{\"u}hne}}, \bibinfo {author} {\bibfnamefont {F.}~\bibnamefont {Paolucci}},
  \bibinfo {author} {\bibfnamefont {J.}~\bibnamefont {Popovic}}, \bibinfo
  {author} {\bibfnamefont {P.~M.}\ \bibnamefont {Ostrovsky}}, \bibinfo {author}
  {\bibfnamefont {J.}~\bibnamefont {Maier}}, \ and\ \bibinfo {author}
  {\bibfnamefont {J.~H.}\ \bibnamefont {Smet}},\ }\href@noop {} {\bibfield
  {journal} {\bibinfo  {journal} {Nature Nanotechnology}\ }\textbf {\bibinfo
  {volume} {12}},\ \bibinfo {pages} {895} (\bibinfo {year} {2017})}\BibitemShut
  {NoStop}%
\bibitem [{\citenamefont {Stra{\ss}er}\ \emph {et~al.}(2015)\citenamefont
  {Stra{\ss}er}, \citenamefont {Ludbrook}, \citenamefont {Levy}, \citenamefont
  {Macdonald}, \citenamefont {Burke}, \citenamefont {Wehling}, \citenamefont
  {Kern}, \citenamefont {Damascelli},\ and\ \citenamefont
  {Ast}}]{Bart.Tl.Graphene.long.and.short.range.scatt.Nano.Lett.2015}%
  \BibitemOpen
  \bibfield  {author} {\bibinfo {author} {\bibfnamefont {C.}~\bibnamefont
  {Stra{\ss}er}}, \bibinfo {author} {\bibfnamefont {B.~M.}\ \bibnamefont
  {Ludbrook}}, \bibinfo {author} {\bibfnamefont {G.}~\bibnamefont {Levy}},
  \bibinfo {author} {\bibfnamefont {A.~J.}\ \bibnamefont {Macdonald}}, \bibinfo
  {author} {\bibfnamefont {S.~A.}\ \bibnamefont {Burke}}, \bibinfo {author}
  {\bibfnamefont {T.~O.}\ \bibnamefont {Wehling}}, \bibinfo {author}
  {\bibfnamefont {K.}~\bibnamefont {Kern}}, \bibinfo {author} {\bibfnamefont
  {A.}~\bibnamefont {Damascelli}}, \ and\ \bibinfo {author} {\bibfnamefont
  {C.~R.}\ \bibnamefont {Ast}},\ }\href {\doibase 10.1021/nl504155f} {\bibfield
   {journal} {\bibinfo  {journal} {Nano Letters}\ }\textbf {\bibinfo {volume}
  {15}},\ \bibinfo {pages} {2825} (\bibinfo {year} {2015})}\BibitemShut
  {NoStop}%
\bibitem [{\citenamefont {Li}\ \emph {et~al.}(2017)\citenamefont {Li},
  \citenamefont {Lin}, \citenamefont {Rui}, \citenamefont {Li}, \citenamefont
  {Zhang}, \citenamefont {Kang}, \citenamefont {Zhang}, \citenamefont {Peng},
  \citenamefont {Liu},\ and\ \citenamefont
  {Xu}}]{nitrogen.graphene.intervalley.short.range.scattering.acsnano.2017}%
  \BibitemOpen
  \bibfield  {author} {\bibinfo {author} {\bibfnamefont {J.}~\bibnamefont
  {Li}}, \bibinfo {author} {\bibfnamefont {L.}~\bibnamefont {Lin}}, \bibinfo
  {author} {\bibfnamefont {D.}~\bibnamefont {Rui}}, \bibinfo {author}
  {\bibfnamefont {Q.}~\bibnamefont {Li}}, \bibinfo {author} {\bibfnamefont
  {J.}~\bibnamefont {Zhang}}, \bibinfo {author} {\bibfnamefont
  {N.}~\bibnamefont {Kang}}, \bibinfo {author} {\bibfnamefont {Y.}~\bibnamefont
  {Zhang}}, \bibinfo {author} {\bibfnamefont {H.}~\bibnamefont {Peng}},
  \bibinfo {author} {\bibfnamefont {Z.}~\bibnamefont {Liu}}, \ and\ \bibinfo
  {author} {\bibfnamefont {H.~Q.}\ \bibnamefont {Xu}},\ }\href {\doibase
  10.1021/acsnano.7b00313} {\bibfield  {journal} {\bibinfo  {journal} {ACS
  Nano}\ }\textbf {\bibinfo {volume} {11}},\ \bibinfo {pages} {4641} (\bibinfo
  {year} {2017})}\BibitemShut {NoStop}%
\bibitem [{\citenamefont {Wellnhofer}\ \emph {et~al.}(2019)\citenamefont
  {Wellnhofer}, \citenamefont {Stabile}, \citenamefont {Kochan}, \citenamefont
  {Gmitra}, \citenamefont {Chuang}, \citenamefont {Zhu},\ and\ \citenamefont
  {Fabian}}]{fluorinated.graphene.theory.exp.PhysRevB.2019}%
  \BibitemOpen
  \bibfield  {author} {\bibinfo {author} {\bibfnamefont {S.}~\bibnamefont
  {Wellnhofer}}, \bibinfo {author} {\bibfnamefont {A.}~\bibnamefont {Stabile}},
  \bibinfo {author} {\bibfnamefont {D.}~\bibnamefont {Kochan}}, \bibinfo
  {author} {\bibfnamefont {M.}~\bibnamefont {Gmitra}}, \bibinfo {author}
  {\bibfnamefont {Y.-W.}\ \bibnamefont {Chuang}}, \bibinfo {author}
  {\bibfnamefont {J.}~\bibnamefont {Zhu}}, \ and\ \bibinfo {author}
  {\bibfnamefont {J.}~\bibnamefont {Fabian}},\ }\href {\doibase
  10.1103/PhysRevB.100.035421} {\bibfield  {journal} {\bibinfo  {journal}
  {Phys. Rev. B}\ }\textbf {\bibinfo {volume} {100}},\ \bibinfo {pages}
  {035421} (\bibinfo {year} {2019})}\BibitemShut {NoStop}%
\bibitem [{\citenamefont {Forti}\ \emph {et~al.}(2011)\citenamefont {Forti},
  \citenamefont {Emtsev}, \citenamefont {Coletti}, \citenamefont {Zakharov},
  \citenamefont {Riedl},\ and\ \citenamefont
  {Starke}}]{SiC.PhysRevB.84.125449}%
  \BibitemOpen
  \bibfield  {author} {\bibinfo {author} {\bibfnamefont {S.}~\bibnamefont
  {Forti}}, \bibinfo {author} {\bibfnamefont {K.~V.}\ \bibnamefont {Emtsev}},
  \bibinfo {author} {\bibfnamefont {C.}~\bibnamefont {Coletti}}, \bibinfo
  {author} {\bibfnamefont {A.~A.}\ \bibnamefont {Zakharov}}, \bibinfo {author}
  {\bibfnamefont {C.}~\bibnamefont {Riedl}}, \ and\ \bibinfo {author}
  {\bibfnamefont {U.}~\bibnamefont {Starke}},\ }\href {\doibase
  10.1103/PhysRevB.84.125449} {\bibfield  {journal} {\bibinfo  {journal} {Phys.
  Rev. B}\ }\textbf {\bibinfo {volume} {84}},\ \bibinfo {pages} {125449}
  (\bibinfo {year} {2011})}\BibitemShut {NoStop}%
\bibitem [{\citenamefont {{graphensic company}}()}]{graphensic}%
  \BibitemOpen
  \bibfield  {author} {\bibinfo {author} {\bibnamefont {{graphensic
  company}}},\ }\href {http://graphensic.com/graphene-material/} {}\bibinfo
  {note} {Epitaxial graphene on silicon carbide;
  http://graphensic.com}\BibitemShut {NoStop}%
\bibitem [{\citenamefont {Khademi}\ \emph {et~al.}(2016)\citenamefont
  {Khademi}, \citenamefont {Sajadi}, \citenamefont {Dosanjh}, \citenamefont
  {Bonn}, \citenamefont {Folk}, \citenamefont {St\"ohr}, \citenamefont
  {Starke},\ and\ \citenamefont {Forti}}]{My.Li.Doping.Paper}%
  \BibitemOpen
  \bibfield  {author} {\bibinfo {author} {\bibfnamefont {A.}~\bibnamefont
  {Khademi}}, \bibinfo {author} {\bibfnamefont {E.}~\bibnamefont {Sajadi}},
  \bibinfo {author} {\bibfnamefont {P.}~\bibnamefont {Dosanjh}}, \bibinfo
  {author} {\bibfnamefont {D.~A.}\ \bibnamefont {Bonn}}, \bibinfo {author}
  {\bibfnamefont {J.~A.}\ \bibnamefont {Folk}}, \bibinfo {author}
  {\bibfnamefont {A.}~\bibnamefont {St\"ohr}}, \bibinfo {author} {\bibfnamefont
  {U.}~\bibnamefont {Starke}}, \ and\ \bibinfo {author} {\bibfnamefont
  {S.}~\bibnamefont {Forti}},\ }\href {\doibase 10.1103/PhysRevB.94.201405}
  {\bibfield  {journal} {\bibinfo  {journal} {Phys. Rev. B}\ }\textbf {\bibinfo
  {volume} {94}},\ \bibinfo {pages} {201405} (\bibinfo {year}
  {2016})}\BibitemShut {NoStop}%
\bibitem [{sup()}]{supplemental}%
  \BibitemOpen
  \href@noop {} {}\bibinfo {note} {See SUPPLEMENTARY INFORMATION for for
  photographs of several samples on the stage, conductivity data for other
  samples, calculated contribution to the dephasing rate from electron-electron
  interactions, weak localization curves’ fitting procedure, theoretical
  tight-binding and Dirac modeling, and discussion about lack of
  superconductivity in Li-doped graphene.}\BibitemShut {Stop}%
\bibitem [{\citenamefont {McCann}\ \emph {et~al.}(2006)\citenamefont {McCann},
  \citenamefont {Kechedzhi}, \citenamefont {Fal'ko}, \citenamefont {Suzuura},
  \citenamefont {Ando},\ and\ \citenamefont
  {Altshuler}}]{McCann.PhysRevLett.2006}%
  \BibitemOpen
  \bibfield  {author} {\bibinfo {author} {\bibfnamefont {E.}~\bibnamefont
  {McCann}}, \bibinfo {author} {\bibfnamefont {K.}~\bibnamefont {Kechedzhi}},
  \bibinfo {author} {\bibfnamefont {V.~I.}\ \bibnamefont {Fal'ko}}, \bibinfo
  {author} {\bibfnamefont {H.}~\bibnamefont {Suzuura}}, \bibinfo {author}
  {\bibfnamefont {T.}~\bibnamefont {Ando}}, \ and\ \bibinfo {author}
  {\bibfnamefont {B.~L.}\ \bibnamefont {Altshuler}},\ }\href {\doibase
  10.1103/PhysRevLett.97.146805} {\bibfield  {journal} {\bibinfo  {journal}
  {Phys. Rev. Lett.}\ }\textbf {\bibinfo {volume} {97}},\ \bibinfo {pages}
  {146805} (\bibinfo {year} {2006})}\BibitemShut {NoStop}%
\bibitem [{\citenamefont {Chen}\ \emph {et~al.}(2010)\citenamefont {Chen},
  \citenamefont {Bae}, \citenamefont {Chialvo}, \citenamefont {Dirks},
  \citenamefont {Bezryadin},\ and\ \citenamefont
  {Mason}}]{intervalley.J.Phys.:Condens.Matter2010}%
  \BibitemOpen
  \bibfield  {author} {\bibinfo {author} {\bibfnamefont {Y.-F.}\ \bibnamefont
  {Chen}}, \bibinfo {author} {\bibfnamefont {M.-H.}\ \bibnamefont {Bae}},
  \bibinfo {author} {\bibfnamefont {C.}~\bibnamefont {Chialvo}}, \bibinfo
  {author} {\bibfnamefont {T.}~\bibnamefont {Dirks}}, \bibinfo {author}
  {\bibfnamefont {A.}~\bibnamefont {Bezryadin}}, \ and\ \bibinfo {author}
  {\bibfnamefont {N.}~\bibnamefont {Mason}},\ }\href
  {http://stacks.iop.org/0953-8984/22/i=20/a=205301} {\bibfield  {journal}
  {\bibinfo  {journal} {Journal of Physics: Condensed Matter}\ }\textbf
  {\bibinfo {volume} {22}},\ \bibinfo {pages} {205301} (\bibinfo {year}
  {2010})}\BibitemShut {NoStop}%
\bibitem [{\citenamefont {Tikhonenko}\ \emph {et~al.}(2008)\citenamefont
  {Tikhonenko}, \citenamefont {Horsell}, \citenamefont {Gorbachev},\ and\
  \citenamefont {Savchenko}}]{wl_gorbachev}%
  \BibitemOpen
  \bibfield  {author} {\bibinfo {author} {\bibfnamefont {F.~V.}\ \bibnamefont
  {Tikhonenko}}, \bibinfo {author} {\bibfnamefont {D.~W.}\ \bibnamefont
  {Horsell}}, \bibinfo {author} {\bibfnamefont {R.~V.}\ \bibnamefont
  {Gorbachev}}, \ and\ \bibinfo {author} {\bibfnamefont {A.~K.}\ \bibnamefont
  {Savchenko}},\ }\href {\doibase 10.1103/PhysRevLett.100.056802} {\bibfield
  {journal} {\bibinfo  {journal} {Phys. Rev. Lett.}\ }\textbf {\bibinfo
  {volume} {100}},\ \bibinfo {pages} {056802} (\bibinfo {year}
  {2008})}\BibitemShut {NoStop}%
\bibitem [{\citenamefont {Lara-Avila}\ \emph {et~al.}(2015)\citenamefont
  {Lara-Avila}, \citenamefont {Kubatkin}, \citenamefont {Kashuba},
  \citenamefont {Folk}, \citenamefont {L\"uscher}, \citenamefont {Yakimova},
  \citenamefont {Janssen}, \citenamefont {Tzalenchuk},\ and\ \citenamefont
  {Fal'ko}}]{PhysRevLett.115.106602}%
  \BibitemOpen
  \bibfield  {author} {\bibinfo {author} {\bibfnamefont {S.}~\bibnamefont
  {Lara-Avila}}, \bibinfo {author} {\bibfnamefont {S.}~\bibnamefont
  {Kubatkin}}, \bibinfo {author} {\bibfnamefont {O.}~\bibnamefont {Kashuba}},
  \bibinfo {author} {\bibfnamefont {J.~A.}\ \bibnamefont {Folk}}, \bibinfo
  {author} {\bibfnamefont {S.}~\bibnamefont {L\"uscher}}, \bibinfo {author}
  {\bibfnamefont {R.}~\bibnamefont {Yakimova}}, \bibinfo {author}
  {\bibfnamefont {T.~J. B.~M.}\ \bibnamefont {Janssen}}, \bibinfo {author}
  {\bibfnamefont {A.}~\bibnamefont {Tzalenchuk}}, \ and\ \bibinfo {author}
  {\bibfnamefont {V.}~\bibnamefont {Fal'ko}},\ }\href {\doibase
  10.1103/PhysRevLett.115.106602} {\bibfield  {journal} {\bibinfo  {journal}
  {Phys. Rev. Lett.}\ }\textbf {\bibinfo {volume} {115}},\ \bibinfo {pages}
  {106602} (\bibinfo {year} {2015})}\BibitemShut {NoStop}%
\bibitem [{\citenamefont {Lara-Avila}\ \emph {et~al.}(2011)\citenamefont
  {Lara-Avila}, \citenamefont {Tzalenchuk}, \citenamefont {Kubatkin},
  \citenamefont {Yakimova}, \citenamefont {Janssen}, \citenamefont {Cedergren},
  \citenamefont {Bergsten},\ and\ \citenamefont
  {Fal'ko}}]{PhysRevLett.107.166602}%
  \BibitemOpen
  \bibfield  {author} {\bibinfo {author} {\bibfnamefont {S.}~\bibnamefont
  {Lara-Avila}}, \bibinfo {author} {\bibfnamefont {A.}~\bibnamefont
  {Tzalenchuk}}, \bibinfo {author} {\bibfnamefont {S.}~\bibnamefont
  {Kubatkin}}, \bibinfo {author} {\bibfnamefont {R.}~\bibnamefont {Yakimova}},
  \bibinfo {author} {\bibfnamefont {T.~J. B.~M.}\ \bibnamefont {Janssen}},
  \bibinfo {author} {\bibfnamefont {K.}~\bibnamefont {Cedergren}}, \bibinfo
  {author} {\bibfnamefont {T.}~\bibnamefont {Bergsten}}, \ and\ \bibinfo
  {author} {\bibfnamefont {V.}~\bibnamefont {Fal'ko}},\ }\href {\doibase
  10.1103/PhysRevLett.107.166602} {\bibfield  {journal} {\bibinfo  {journal}
  {Phys. Rev. Lett.}\ }\textbf {\bibinfo {volume} {107}},\ \bibinfo {pages}
  {166602} (\bibinfo {year} {2011})}\BibitemShut {NoStop}%
\bibitem [{\citenamefont {Yan}\ \emph {et~al.}(2016)\citenamefont {Yan},
  \citenamefont {Han}, \citenamefont {Jia}, \citenamefont {Niu}, \citenamefont
  {Cai}, \citenamefont {Yu},\ and\ \citenamefont
  {Wu}}]{Intervalley.charge.state.of.defects.PhysRevB.2016}%
  \BibitemOpen
  \bibfield  {author} {\bibinfo {author} {\bibfnamefont {B.}~\bibnamefont
  {Yan}}, \bibinfo {author} {\bibfnamefont {Q.}~\bibnamefont {Han}}, \bibinfo
  {author} {\bibfnamefont {Z.}~\bibnamefont {Jia}}, \bibinfo {author}
  {\bibfnamefont {J.}~\bibnamefont {Niu}}, \bibinfo {author} {\bibfnamefont
  {T.}~\bibnamefont {Cai}}, \bibinfo {author} {\bibfnamefont {D.}~\bibnamefont
  {Yu}}, \ and\ \bibinfo {author} {\bibfnamefont {X.}~\bibnamefont {Wu}},\
  }\href {\doibase 10.1103/PhysRevB.93.041407} {\bibfield  {journal} {\bibinfo
  {journal} {Phys. Rev. B}\ }\textbf {\bibinfo {volume} {93}},\ \bibinfo
  {pages} {041407} (\bibinfo {year} {2016})}\BibitemShut {NoStop}%
\bibitem [{\citenamefont {Mallet}\ \emph {et~al.}(2012)\citenamefont {Mallet},
  \citenamefont {Brihuega}, \citenamefont {Bose}, \citenamefont {Ugeda},
  \citenamefont {G\'omez-Rodr\'{\i}guez}, \citenamefont {Kern},\ and\
  \citenamefont {Veuillen}}]{SR.disorder.SiC.graphene.Phys.Rev.B.2012}%
  \BibitemOpen
  \bibfield  {author} {\bibinfo {author} {\bibfnamefont {P.}~\bibnamefont
  {Mallet}}, \bibinfo {author} {\bibfnamefont {I.}~\bibnamefont {Brihuega}},
  \bibinfo {author} {\bibfnamefont {S.}~\bibnamefont {Bose}}, \bibinfo {author}
  {\bibfnamefont {M.~M.}\ \bibnamefont {Ugeda}}, \bibinfo {author}
  {\bibfnamefont {J.~M.}\ \bibnamefont {G\'omez-Rodr\'{\i}guez}}, \bibinfo
  {author} {\bibfnamefont {K.}~\bibnamefont {Kern}}, \ and\ \bibinfo {author}
  {\bibfnamefont {J.~Y.}\ \bibnamefont {Veuillen}},\ }\href {\doibase
  10.1103/PhysRevB.86.045444} {\bibfield  {journal} {\bibinfo  {journal} {Phys.
  Rev. B}\ }\textbf {\bibinfo {volume} {86}},\ \bibinfo {pages} {045444}
  (\bibinfo {year} {2012})}\BibitemShut {NoStop}%
\bibitem [{\citenamefont {Fan}\ \emph {et~al.}(2013)\citenamefont {Fan},
  \citenamefont {Zheng}, \citenamefont {Kuo},\ and\ \citenamefont
  {Singh}}]{Li.Cluster.ACS.Appl.Mater.Interfaces.2013}%
  \BibitemOpen
  \bibfield  {author} {\bibinfo {author} {\bibfnamefont {X.}~\bibnamefont
  {Fan}}, \bibinfo {author} {\bibfnamefont {W.~T.}\ \bibnamefont {Zheng}},
  \bibinfo {author} {\bibfnamefont {J.-L.}\ \bibnamefont {Kuo}}, \ and\
  \bibinfo {author} {\bibfnamefont {D.~J.}\ \bibnamefont {Singh}},\ }\href
  {\doibase 10.1021/am401548c} {\bibfield  {journal} {\bibinfo  {journal} {ACS
  Applied Materials \& Interfaces}\ }\textbf {\bibinfo {volume} {5}},\ \bibinfo
  {pages} {7793} (\bibinfo {year} {2013})},\ \bibinfo {note} {pMID:
  23863039}\BibitemShut {NoStop}%
\bibitem [{\citenamefont {Liu}\ \emph {et~al.}(2014)\citenamefont {Liu},
  \citenamefont {Kutana}, \citenamefont {Liu},\ and\ \citenamefont
  {Yakobson}}]{Li.Cluster.J.Phys.Chem.Lett.2014}%
  \BibitemOpen
  \bibfield  {author} {\bibinfo {author} {\bibfnamefont {M.}~\bibnamefont
  {Liu}}, \bibinfo {author} {\bibfnamefont {A.}~\bibnamefont {Kutana}},
  \bibinfo {author} {\bibfnamefont {Y.}~\bibnamefont {Liu}}, \ and\ \bibinfo
  {author} {\bibfnamefont {B.~I.}\ \bibnamefont {Yakobson}},\ }\href {\doibase
  10.1021/jz500199d} {\bibfield  {journal} {\bibinfo  {journal} {The Journal of
  Physical Chemistry Letters}\ }\textbf {\bibinfo {volume} {5}},\ \bibinfo
  {pages} {1225} (\bibinfo {year} {2014})},\ \bibinfo {note} {pMID:
  26274475}\BibitemShut {NoStop}%
\bibitem [{\citenamefont {Valencia}\ \emph {et~al.}(2006)\citenamefont
  {Valencia}, \citenamefont {Romero}, \citenamefont {Ancilotto},\ and\
  \citenamefont {Silvestrelli}}]{DFT.J.Phys.Chem.B2006}%
  \BibitemOpen
  \bibfield  {author} {\bibinfo {author} {\bibfnamefont {F.}~\bibnamefont
  {Valencia}}, \bibinfo {author} {\bibfnamefont {A.~H.}\ \bibnamefont
  {Romero}}, \bibinfo {author} {\bibfnamefont {F.}~\bibnamefont {Ancilotto}}, \
  and\ \bibinfo {author} {\bibfnamefont {P.~L.}\ \bibnamefont {Silvestrelli}},\
  }\href {\doibase 10.1021/jp062126+} {\bibfield  {journal} {\bibinfo
  {journal} {The Journal of Physical Chemistry B}\ }\textbf {\bibinfo {volume}
  {110}},\ \bibinfo {pages} {14832} (\bibinfo {year} {2006})}\BibitemShut
  {NoStop}%
\bibitem [{\citenamefont {Farjam}\ and\ \citenamefont
  {Rafii-Tabar}(2009)}]{DFT.Li.Graphene.Gap.PhysRevB.2009}%
  \BibitemOpen
  \bibfield  {author} {\bibinfo {author} {\bibfnamefont {M.}~\bibnamefont
  {Farjam}}\ and\ \bibinfo {author} {\bibfnamefont {H.}~\bibnamefont
  {Rafii-Tabar}},\ }\href {\doibase 10.1103/PhysRevB.79.045417} {\bibfield
  {journal} {\bibinfo  {journal} {Phys. Rev. B}\ }\textbf {\bibinfo {volume}
  {79}},\ \bibinfo {pages} {045417} (\bibinfo {year} {2009})}\BibitemShut
  {NoStop}%
\bibitem [{\citenamefont {Wehling}\ \emph {et~al.}(2009)\citenamefont
  {Wehling}, \citenamefont {Katsnelson},\ and\ \citenamefont
  {Lichtenstein}}]{Wehling.res.PhysRevB.2009}%
  \BibitemOpen
  \bibfield  {author} {\bibinfo {author} {\bibfnamefont {T.~O.}\ \bibnamefont
  {Wehling}}, \bibinfo {author} {\bibfnamefont {M.~I.}\ \bibnamefont
  {Katsnelson}}, \ and\ \bibinfo {author} {\bibfnamefont {A.~I.}\ \bibnamefont
  {Lichtenstein}},\ }\href {\doibase 10.1103/PhysRevB.80.085428} {\bibfield
  {journal} {\bibinfo  {journal} {Phys. Rev. B}\ }\textbf {\bibinfo {volume}
  {80}},\ \bibinfo {pages} {085428} (\bibinfo {year} {2009})}\BibitemShut
  {NoStop}%
\bibitem [{\citenamefont {Wehling}\ \emph {et~al.}(2010)\citenamefont
  {Wehling}, \citenamefont {Yuan}, \citenamefont {Lichtenstein}, \citenamefont
  {Geim},\ and\ \citenamefont {Katsnelson}}]{Wehling.res.PhysRevLett.2010}%
  \BibitemOpen
  \bibfield  {author} {\bibinfo {author} {\bibfnamefont {T.~O.}\ \bibnamefont
  {Wehling}}, \bibinfo {author} {\bibfnamefont {S.}~\bibnamefont {Yuan}},
  \bibinfo {author} {\bibfnamefont {A.~I.}\ \bibnamefont {Lichtenstein}},
  \bibinfo {author} {\bibfnamefont {A.~K.}\ \bibnamefont {Geim}}, \ and\
  \bibinfo {author} {\bibfnamefont {M.~I.}\ \bibnamefont {Katsnelson}},\ }\href
  {\doibase 10.1103/PhysRevLett.105.056802} {\bibfield  {journal} {\bibinfo
  {journal} {Phys. Rev. Lett.}\ }\textbf {\bibinfo {volume} {105}},\ \bibinfo
  {pages} {056802} (\bibinfo {year} {2010})}\BibitemShut {NoStop}%
\bibitem [{\citenamefont {Irmer}\ \emph {et~al.}(2018)\citenamefont {Irmer},
  \citenamefont {Kochan}, \citenamefont {Lee},\ and\ \citenamefont
  {Fabian}}]{Irmer.PhysRevB.2018}%
  \BibitemOpen
  \bibfield  {author} {\bibinfo {author} {\bibfnamefont {S.}~\bibnamefont
  {Irmer}}, \bibinfo {author} {\bibfnamefont {D.}~\bibnamefont {Kochan}},
  \bibinfo {author} {\bibfnamefont {J.}~\bibnamefont {Lee}}, \ and\ \bibinfo
  {author} {\bibfnamefont {J.}~\bibnamefont {Fabian}},\ }\href {\doibase
  10.1103/PhysRevB.97.075417} {\bibfield  {journal} {\bibinfo  {journal} {Phys.
  Rev. B}\ }\textbf {\bibinfo {volume} {97}},\ \bibinfo {pages} {075417}
  (\bibinfo {year} {2018})}\BibitemShut {NoStop}%
\bibitem [{\citenamefont {Fowler}\ and\ \citenamefont
  {Hartstein}(1980)}]{Hartstein.Fowler.1980}%
  \BibitemOpen
  \bibfield  {author} {\bibinfo {author} {\bibfnamefont {A.~B.}\ \bibnamefont
  {Fowler}}\ and\ \bibinfo {author} {\bibfnamefont {A.}~\bibnamefont
  {Hartstein}},\ }\href {\doibase 10.1080/01418638008222339} {\bibfield
  {journal} {\bibinfo  {journal} {Philosophical Magazine B}\ }\textbf {\bibinfo
  {volume} {42}},\ \bibinfo {pages} {949} (\bibinfo {year} {1980})}\BibitemShut
  {NoStop}%
\bibitem [{\citenamefont {Ando}\ \emph {et~al.}(1982)\citenamefont {Ando},
  \citenamefont {Fowler},\ and\ \citenamefont {Stern}}]{RevModPhys.1982}%
  \BibitemOpen
  \bibfield  {author} {\bibinfo {author} {\bibfnamefont {T.}~\bibnamefont
  {Ando}}, \bibinfo {author} {\bibfnamefont {A.~B.}\ \bibnamefont {Fowler}}, \
  and\ \bibinfo {author} {\bibfnamefont {F.}~\bibnamefont {Stern}},\ }\href
  {\doibase 10.1103/RevModPhys.54.437} {\bibfield  {journal} {\bibinfo
  {journal} {Rev. Mod. Phys.}\ }\textbf {\bibinfo {volume} {54}},\ \bibinfo
  {pages} {437} (\bibinfo {year} {1982})}\BibitemShut {NoStop}%
\bibitem [{\citenamefont {Boross}\ and\ \citenamefont
  {P\'alyi}(2015)}]{Intervalley.Charged.Scattering.PhysRevB.2015}%
  \BibitemOpen
  \bibfield  {author} {\bibinfo {author} {\bibfnamefont {P.}~\bibnamefont
  {Boross}}\ and\ \bibinfo {author} {\bibfnamefont {A.}~\bibnamefont
  {P\'alyi}},\ }\href {\doibase 10.1103/PhysRevB.92.035420} {\bibfield
  {journal} {\bibinfo  {journal} {Phys. Rev. B}\ }\textbf {\bibinfo {volume}
  {92}},\ \bibinfo {pages} {035420} (\bibinfo {year} {2015})}\BibitemShut
  {NoStop}%
\bibitem [{\citenamefont {Stauber}\ \emph {et~al.}(2017)\citenamefont
  {Stauber}, \citenamefont {Parida}, \citenamefont {Trushin}, \citenamefont
  {Ulybyshev}, \citenamefont {Boyda},\ and\ \citenamefont
  {Schliemann}}]{Schliemann:Interacting}%
  \BibitemOpen
  \bibfield  {author} {\bibinfo {author} {\bibfnamefont {T.}~\bibnamefont
  {Stauber}}, \bibinfo {author} {\bibfnamefont {P.}~\bibnamefont {Parida}},
  \bibinfo {author} {\bibfnamefont {M.}~\bibnamefont {Trushin}}, \bibinfo
  {author} {\bibfnamefont {M.~V.}\ \bibnamefont {Ulybyshev}}, \bibinfo {author}
  {\bibfnamefont {D.~L.}\ \bibnamefont {Boyda}}, \ and\ \bibinfo {author}
  {\bibfnamefont {J.}~\bibnamefont {Schliemann}},\ }\href {\doibase
  10.1103/PhysRevLett.118.266801} {\bibfield  {journal} {\bibinfo  {journal}
  {Phys. Rev. Lett.}\ }\textbf {\bibinfo {volume} {118}},\ \bibinfo {pages}
  {266801} (\bibinfo {year} {2017})}\BibitemShut {NoStop}%
\bibitem [{\citenamefont {Chandni}\ \emph {et~al.}(2015)\citenamefont
  {Chandni}, \citenamefont {Henriksen},\ and\ \citenamefont
  {Eisenstein}}]{In.12K.PhysRevB.2015}%
  \BibitemOpen
  \bibfield  {author} {\bibinfo {author} {\bibfnamefont {U.}~\bibnamefont
  {Chandni}}, \bibinfo {author} {\bibfnamefont {E.~A.}\ \bibnamefont
  {Henriksen}}, \ and\ \bibinfo {author} {\bibfnamefont {J.~P.}\ \bibnamefont
  {Eisenstein}},\ }\href {\doibase 10.1103/PhysRevB.91.245402} {\bibfield
  {journal} {\bibinfo  {journal} {Phys. Rev. B}\ }\textbf {\bibinfo {volume}
  {91}},\ \bibinfo {pages} {245402} (\bibinfo {year} {2015})}\BibitemShut
  {NoStop}%
\end{thebibliography}%


\begin{thebibliography}{23}%
\makeatletter
\providecommand \@ifxundefined [1]{%
 \@ifx{#1\undefined}
}%
\providecommand \@ifnum [1]{%
 \ifnum #1\expandafter \@firstoftwo
 \else \expandafter \@secondoftwo
 \fi
}%
\providecommand \@ifx [1]{%
 \ifx #1\expandafter \@firstoftwo
 \else \expandafter \@secondoftwo
 \fi
}%
\providecommand \natexlab [1]{#1}%
\providecommand \enquote  [1]{``#1''}%
\providecommand \bibnamefont  [1]{#1}%
\providecommand \bibfnamefont [1]{#1}%
\providecommand \citenamefont [1]{#1}%
\providecommand \href@noop [0]{\@secondoftwo}%
\providecommand \href [0]{\begingroup \@sanitize@url \@href}%
\providecommand \@href[1]{\@@startlink{#1}\@@href}%
\providecommand \@@href[1]{\endgroup#1\@@endlink}%
\providecommand \@sanitize@url [0]{\catcode `\\12\catcode `\$12\catcode
  `\&12\catcode `\#12\catcode `\^12\catcode `\_12\catcode `\%12\relax}%
\providecommand \@@startlink[1]{}%
\providecommand \@@endlink[0]{}%
\providecommand \url  [0]{\begingroup\@sanitize@url \@url }%
\providecommand \@url [1]{\endgroup\@href {#1}{\urlprefix }}%
\providecommand \urlprefix  [0]{URL }%
\providecommand \Eprint [0]{\href }%
\providecommand \doibase [0]{http://dx.doi.org/}%
\providecommand \selectlanguage [0]{\@gobble}%
\providecommand \bibinfo  [0]{\@secondoftwo}%
\providecommand \bibfield  [0]{\@secondoftwo}%
\providecommand \translation [1]{[#1]}%
\providecommand \BibitemOpen [0]{}%
\providecommand \bibitemStop [0]{}%
\providecommand \bibitemNoStop [0]{.\EOS\space}%
\providecommand \EOS [0]{\spacefactor3000\relax}%
\providecommand \BibitemShut  [1]{\csname bibitem#1\endcsname}%
\let\auto@bib@innerbib\@empty
\bibitem [{\citenamefont {Khademi}\ \emph {et~al.}(2016)\citenamefont
  {Khademi}, \citenamefont {Sajadi}, \citenamefont {Dosanjh}, \citenamefont
  {Bonn}, \citenamefont {Folk}, \citenamefont {St\"ohr}, \citenamefont
  {Starke},\ and\ \citenamefont {Forti}}]{s.My.Li.Doping.Paper}%
  \BibitemOpen
  \bibfield  {author} {\bibinfo {author} {\bibfnamefont {A.}~\bibnamefont
  {Khademi}}, \bibinfo {author} {\bibfnamefont {E.}~\bibnamefont {Sajadi}},
  \bibinfo {author} {\bibfnamefont {P.}~\bibnamefont {Dosanjh}}, \bibinfo
  {author} {\bibfnamefont {D.~A.}\ \bibnamefont {Bonn}}, \bibinfo {author}
  {\bibfnamefont {J.~A.}\ \bibnamefont {Folk}}, \bibinfo {author}
  {\bibfnamefont {A.}~\bibnamefont {St\"ohr}}, \bibinfo {author} {\bibfnamefont
  {U.}~\bibnamefont {Starke}}, \ and\ \bibinfo {author} {\bibfnamefont
  {S.}~\bibnamefont {Forti}},\ }\href {\doibase 10.1103/PhysRevB.94.201405}
  {\bibfield  {journal} {\bibinfo  {journal} {Phys. Rev. B}\ }\textbf {\bibinfo
  {volume} {94}},\ \bibinfo {pages} {201405} (\bibinfo {year}
  {2016})}\BibitemShut {NoStop}%
\bibitem [{\citenamefont {Altshuler}\ \emph {et~al.}(1980)\citenamefont
  {Altshuler}, \citenamefont {Khmel'nitzkii}, \citenamefont {Larkin},\ and\
  \citenamefont {Lee}}]{s.Altshuler.PhysRevB.1980}%
  \BibitemOpen
  \bibfield  {author} {\bibinfo {author} {\bibfnamefont {B.~L.}\ \bibnamefont
  {Altshuler}}, \bibinfo {author} {\bibfnamefont {D.}~\bibnamefont
  {Khmel'nitzkii}}, \bibinfo {author} {\bibfnamefont {A.~I.}\ \bibnamefont
  {Larkin}}, \ and\ \bibinfo {author} {\bibfnamefont {P.~A.}\ \bibnamefont
  {Lee}},\ }\href {\doibase 10.1103/PhysRevB.22.5142} {\bibfield  {journal}
  {\bibinfo  {journal} {Phys. Rev. B}\ }\textbf {\bibinfo {volume} {22}},\
  \bibinfo {pages} {5142} (\bibinfo {year} {1980})}\BibitemShut {NoStop}%
\bibitem [{\citenamefont {Lara-Avila}\ \emph {et~al.}(2015)\citenamefont
  {Lara-Avila}, \citenamefont {Kubatkin}, \citenamefont {Kashuba},
  \citenamefont {Folk}, \citenamefont {L\"uscher}, \citenamefont {Yakimova},
  \citenamefont {Janssen}, \citenamefont {Tzalenchuk},\ and\ \citenamefont
  {Fal'ko}}]{s.PhysRevLett.115.106602}%
  \BibitemOpen
  \bibfield  {author} {\bibinfo {author} {\bibfnamefont {S.}~\bibnamefont
  {Lara-Avila}}, \bibinfo {author} {\bibfnamefont {S.}~\bibnamefont
  {Kubatkin}}, \bibinfo {author} {\bibfnamefont {O.}~\bibnamefont {Kashuba}},
  \bibinfo {author} {\bibfnamefont {J.~A.}\ \bibnamefont {Folk}}, \bibinfo
  {author} {\bibfnamefont {S.}~\bibnamefont {L\"uscher}}, \bibinfo {author}
  {\bibfnamefont {R.}~\bibnamefont {Yakimova}}, \bibinfo {author}
  {\bibfnamefont {T.~J. B.~M.}\ \bibnamefont {Janssen}}, \bibinfo {author}
  {\bibfnamefont {A.}~\bibnamefont {Tzalenchuk}}, \ and\ \bibinfo {author}
  {\bibfnamefont {V.}~\bibnamefont {Fal'ko}},\ }\href {\doibase
  10.1103/PhysRevLett.115.106602} {\bibfield  {journal} {\bibinfo  {journal}
  {Phys. Rev. Lett.}\ }\textbf {\bibinfo {volume} {115}},\ \bibinfo {pages}
  {106602} (\bibinfo {year} {2015})}\BibitemShut {NoStop}%
\bibitem [{\citenamefont {Bevington}(1969)}]{s.chi.square.Book}%
  \BibitemOpen
  \bibfield  {author} {\bibinfo {author} {\bibfnamefont {P.~R.}\ \bibnamefont
  {Bevington}},\ }\href@noop {} {\emph {\bibinfo {title} {Data Reduction and
  Error Analysis for the Physical Sciences}}}\ (\bibinfo  {publisher}
  {McGraw-Hill},\ \bibinfo {address} {New York},\ \bibinfo {year}
  {1969})\BibitemShut {NoStop}%
\bibitem [{\citenamefont {Hwang}\ \emph {et~al.}(2007)\citenamefont {Hwang},
  \citenamefont {Adam},\ and\ \citenamefont {Sarma}}]{s.Sarma:Carrier}%
  \BibitemOpen
  \bibfield  {author} {\bibinfo {author} {\bibfnamefont {E.~H.}\ \bibnamefont
  {Hwang}}, \bibinfo {author} {\bibfnamefont {S.}~\bibnamefont {Adam}}, \ and\
  \bibinfo {author} {\bibfnamefont {S.~D.}\ \bibnamefont {Sarma}},\ }\href@noop
  {} {\bibfield  {journal} {\bibinfo  {journal} {Phys. Rev. Lett.}\ }\textbf
  {\bibinfo {volume} {98}},\ \bibinfo {pages} {186806} (\bibinfo {year}
  {2007})}\BibitemShut {NoStop}%
\bibitem [{\citenamefont {Stauber}\ \emph {et~al.}(2017)\citenamefont
  {Stauber}, \citenamefont {Parida}, \citenamefont {Trushin}, \citenamefont
  {Ulybyshev}, \citenamefont {Boyda},\ and\ \citenamefont
  {Schliemann}}]{Schliemann:Interacting}%
  \BibitemOpen
  \bibfield  {author} {\bibinfo {author} {\bibfnamefont {T.}~\bibnamefont
  {Stauber}}, \bibinfo {author} {\bibfnamefont {P.}~\bibnamefont {Parida}},
  \bibinfo {author} {\bibfnamefont {M.}~\bibnamefont {Trushin}}, \bibinfo
  {author} {\bibfnamefont {M.~V.}\ \bibnamefont {Ulybyshev}}, \bibinfo {author}
  {\bibfnamefont {D.~L.}\ \bibnamefont {Boyda}}, \ and\ \bibinfo {author}
  {\bibfnamefont {J.}~\bibnamefont {Schliemann}},\ }\href@noop {} {\bibfield
  {journal} {\bibinfo  {journal} {Phys. Rev. Lett.}\ }\textbf {\bibinfo
  {volume} {118}},\ \bibinfo {pages} {266801} (\bibinfo {year}
  {2017})}\BibitemShut {NoStop}%
\bibitem [{\citenamefont {Kaasbjerg}\ and\ \citenamefont
  {Jauho}(2019)}]{s.Kristen}%
  \BibitemOpen
  \bibfield  {author} {\bibinfo {author} {\bibfnamefont {K.}~\bibnamefont
  {Kaasbjerg}}\ and\ \bibinfo {author} {\bibfnamefont {A.-P.}\ \bibnamefont
  {Jauho}},\ }\href@noop {} {\bibfield  {journal} {\bibinfo  {journal} {Phys.
  Rev. XXX}\ }\textbf {\bibinfo {volume} {XXX}},\ \bibinfo {pages} {XXX}
  (\bibinfo {year} {2019})},\ \bibinfo {note} {arXiv:1904.08191}\BibitemShut
  {NoStop}%
\bibitem [{\citenamefont {Adam}\ \emph {et~al.}(2007)\citenamefont {Adam},
  \citenamefont {Hwang}, \citenamefont {Galitski},\ and\ \citenamefont
  {Das~Sarma}}]{sAdamPNAS2007}%
  \BibitemOpen
  \bibfield  {author} {\bibinfo {author} {\bibfnamefont {S.}~\bibnamefont
  {Adam}}, \bibinfo {author} {\bibfnamefont {E.~H.}\ \bibnamefont {Hwang}},
  \bibinfo {author} {\bibfnamefont {V.~M.}\ \bibnamefont {Galitski}}, \ and\
  \bibinfo {author} {\bibfnamefont {S.}~\bibnamefont {Das~Sarma}},\ }\href
  {\doibase 10.1073/pnas.0704772104} {\bibfield  {journal} {\bibinfo  {journal}
  {Proc. Natl. Acad. Sci. U.S.A.}\ }\textbf {\bibinfo {volume} {104}},\
  \bibinfo {pages} {18392} (\bibinfo {year} {2007})}\BibitemShut {NoStop}%
\bibitem [{\citenamefont {Hwang}\ \emph {et~al.}(2012)\citenamefont {Hwang},
  \citenamefont {Siegel}, \citenamefont {Mo}, \citenamefont {Regan},
  \citenamefont {Ismach}, \citenamefont {Zhang}, \citenamefont {Zettl},\ and\
  \citenamefont {Lanzara}}]{s.SiC.dielectric.constant.Sci.Rep.2012}%
  \BibitemOpen
  \bibfield  {author} {\bibinfo {author} {\bibfnamefont {C.}~\bibnamefont
  {Hwang}}, \bibinfo {author} {\bibfnamefont {D.~A.}\ \bibnamefont {Siegel}},
  \bibinfo {author} {\bibfnamefont {S.-K.}\ \bibnamefont {Mo}}, \bibinfo
  {author} {\bibfnamefont {W.}~\bibnamefont {Regan}}, \bibinfo {author}
  {\bibfnamefont {A.}~\bibnamefont {Ismach}}, \bibinfo {author} {\bibfnamefont
  {Y.}~\bibnamefont {Zhang}}, \bibinfo {author} {\bibfnamefont
  {A.}~\bibnamefont {Zettl}}, \ and\ \bibinfo {author} {\bibfnamefont
  {A.}~\bibnamefont {Lanzara}},\ }\href {http://dx.doi.org/10.1038/srep00590}
  {\bibfield  {journal} {\bibinfo  {journal} {Sci Rep.}\ }\textbf {\bibinfo
  {volume} {2}},\ \bibinfo {pages} {590 EP } (\bibinfo {year} {2012})},\
  \bibinfo {note} {article}\BibitemShut {NoStop}%
\bibitem [{\citenamefont {Stra{\ss}er}\ \emph {et~al.}(2015)\citenamefont
  {Stra{\ss}er}, \citenamefont {Ludbrook}, \citenamefont {Levy}, \citenamefont
  {Macdonald}, \citenamefont {Burke}, \citenamefont {Wehling}, \citenamefont
  {Kern}, \citenamefont {Damascelli},\ and\ \citenamefont
  {Ast}}]{s.Bart.Tl.Graphene.long.and.short.range.scatt.Nano.Lett.2015}%
  \BibitemOpen
  \bibfield  {author} {\bibinfo {author} {\bibfnamefont {C.}~\bibnamefont
  {Stra{\ss}er}}, \bibinfo {author} {\bibfnamefont {B.~M.}\ \bibnamefont
  {Ludbrook}}, \bibinfo {author} {\bibfnamefont {G.}~\bibnamefont {Levy}},
  \bibinfo {author} {\bibfnamefont {A.~J.}\ \bibnamefont {Macdonald}}, \bibinfo
  {author} {\bibfnamefont {S.~A.}\ \bibnamefont {Burke}}, \bibinfo {author}
  {\bibfnamefont {T.~O.}\ \bibnamefont {Wehling}}, \bibinfo {author}
  {\bibfnamefont {K.}~\bibnamefont {Kern}}, \bibinfo {author} {\bibfnamefont
  {A.}~\bibnamefont {Damascelli}}, \ and\ \bibinfo {author} {\bibfnamefont
  {C.~R.}\ \bibnamefont {Ast}},\ }\href {\doibase 10.1021/nl504155f} {\bibfield
   {journal} {\bibinfo  {journal} {Nano Lett.}\ }\textbf {\bibinfo {volume}
  {15}},\ \bibinfo {pages} {2825} (\bibinfo {year} {2015})}\BibitemShut
  {NoStop}%
\bibitem [{\citenamefont {Profeta}\ \emph {et~al.}(2012)\citenamefont
  {Profeta}, \citenamefont {Calandra},\ and\ \citenamefont
  {Mauri}}]{s.LiSC.NatPhys2012}%
  \BibitemOpen
  \bibfield  {author} {\bibinfo {author} {\bibfnamefont {G.}~\bibnamefont
  {Profeta}}, \bibinfo {author} {\bibfnamefont {M.}~\bibnamefont {Calandra}}, \
  and\ \bibinfo {author} {\bibfnamefont {F.}~\bibnamefont {Mauri}},\ }\href
  {\doibase 10.1038/nphys2181} {\bibfield  {journal} {\bibinfo  {journal} {Nat.
  Phys.}\ }\textbf {\bibinfo {volume} {8}},\ \bibinfo {pages} {131} (\bibinfo
  {year} {2012})}\BibitemShut {NoStop}%
\bibitem [{\citenamefont {Ludbrook}\ \emph {et~al.}(2015)\citenamefont
  {Ludbrook}, \citenamefont {Levy}, \citenamefont {Nigge}, \citenamefont
  {Zonno}, \citenamefont {Schneider}, \citenamefont {Dvorak}, \citenamefont
  {Veenstra}, \citenamefont {Zhdanovich}, \citenamefont {Wong}, \citenamefont
  {Dosanjh}, \citenamefont {Stra{\ss}er}, \citenamefont {St\"ohr},
  \citenamefont {Forti}, \citenamefont {Ast}, \citenamefont {Starke},\ and\
  \citenamefont {Damascelli}}]{s.Bart}%
  \BibitemOpen
  \bibfield  {author} {\bibinfo {author} {\bibfnamefont {B.~M.}\ \bibnamefont
  {Ludbrook}}, \bibinfo {author} {\bibfnamefont {G.}~\bibnamefont {Levy}},
  \bibinfo {author} {\bibfnamefont {P.}~\bibnamefont {Nigge}}, \bibinfo
  {author} {\bibfnamefont {M.}~\bibnamefont {Zonno}}, \bibinfo {author}
  {\bibfnamefont {M.}~\bibnamefont {Schneider}}, \bibinfo {author}
  {\bibfnamefont {D.~J.}\ \bibnamefont {Dvorak}}, \bibinfo {author}
  {\bibfnamefont {C.~N.}\ \bibnamefont {Veenstra}}, \bibinfo {author}
  {\bibfnamefont {S.}~\bibnamefont {Zhdanovich}}, \bibinfo {author}
  {\bibfnamefont {D.}~\bibnamefont {Wong}}, \bibinfo {author} {\bibfnamefont
  {P.}~\bibnamefont {Dosanjh}}, \bibinfo {author} {\bibfnamefont
  {C.}~\bibnamefont {Stra{\ss}er}}, \bibinfo {author} {\bibfnamefont
  {A.}~\bibnamefont {St\"ohr}}, \bibinfo {author} {\bibfnamefont
  {S.}~\bibnamefont {Forti}}, \bibinfo {author} {\bibfnamefont {C.~R.}\
  \bibnamefont {Ast}}, \bibinfo {author} {\bibfnamefont {U.}~\bibnamefont
  {Starke}}, \ and\ \bibinfo {author} {\bibfnamefont {A.}~\bibnamefont
  {Damascelli}},\ }\href {\doibase 10.1073/pnas.1510435112} {\bibfield
  {journal} {\bibinfo  {journal} {Proc. Natl. Acad. Sci. U.S.A.}\ }\textbf
  {\bibinfo {volume} {112}},\ \bibinfo {pages} {11795} (\bibinfo {year}
  {2015})}\BibitemShut {NoStop}%
\bibitem [{\citenamefont {Ichinokura}\ \emph {et~al.}(2016)\citenamefont
  {Ichinokura}, \citenamefont {Sugawara}, \citenamefont {Takayama},
  \citenamefont {Takahashi},\ and\ \citenamefont
  {Hasegawa}}]{s.SC.Ca.bilayer.graphene.ACSNano.2016}%
  \BibitemOpen
  \bibfield  {author} {\bibinfo {author} {\bibfnamefont {S.}~\bibnamefont
  {Ichinokura}}, \bibinfo {author} {\bibfnamefont {K.}~\bibnamefont
  {Sugawara}}, \bibinfo {author} {\bibfnamefont {A.}~\bibnamefont {Takayama}},
  \bibinfo {author} {\bibfnamefont {T.}~\bibnamefont {Takahashi}}, \ and\
  \bibinfo {author} {\bibfnamefont {S.}~\bibnamefont {Hasegawa}},\ }\href
  {\doibase 10.1021/acsnano.5b07848} {\bibfield  {journal} {\bibinfo  {journal}
  {ACS Nano}\ }\textbf {\bibinfo {volume} {10}},\ \bibinfo {pages} {2761}
  (\bibinfo {year} {2016})}\BibitemShut {NoStop}%
\bibitem [{\citenamefont
  {Patton}(1971)}]{s.2D.superconducting.fluctuations.PhysRevLett.1971}%
  \BibitemOpen
  \bibfield  {author} {\bibinfo {author} {\bibfnamefont {B.~R.}\ \bibnamefont
  {Patton}},\ }\href {\doibase 10.1103/PhysRevLett.27.1273} {\bibfield
  {journal} {\bibinfo  {journal} {Phys. Rev. Lett.}\ }\textbf {\bibinfo
  {volume} {27}},\ \bibinfo {pages} {1273} (\bibinfo {year}
  {1971})}\BibitemShut {NoStop}%
\bibitem [{\citenamefont {Aslamasov}\ and\ \citenamefont
  {Larkin}(1968)}]{s.low.D.SC.Phys.Lett.A.1968}%
  \BibitemOpen
  \bibfield  {author} {\bibinfo {author} {\bibfnamefont {L.}~\bibnamefont
  {Aslamasov}}\ and\ \bibinfo {author} {\bibfnamefont {A.}~\bibnamefont
  {Larkin}},\ }\href {\doibase http://dx.doi.org/10.1016/0375-9601(68)90623-3}
  {\bibfield  {journal} {\bibinfo  {journal} {Phys. Lett. A}\ }\textbf
  {\bibinfo {volume} {26}},\ \bibinfo {pages} {238 } (\bibinfo {year}
  {1968})}\BibitemShut {NoStop}%
\bibitem [{\citenamefont {Hetel}\ \emph {et~al.}(2007)\citenamefont {Hetel},
  \citenamefont {Lemberger},\ and\ \citenamefont
  {Randeria}}]{s.BKT.Nat.Phys.2007}%
  \BibitemOpen
  \bibfield  {author} {\bibinfo {author} {\bibfnamefont {I.}~\bibnamefont
  {Hetel}}, \bibinfo {author} {\bibfnamefont {T.~R.}\ \bibnamefont
  {Lemberger}}, \ and\ \bibinfo {author} {\bibfnamefont {M.}~\bibnamefont
  {Randeria}},\ }\href {\doibase 10.1038/nphys707} {\bibfield  {journal}
  {\bibinfo  {journal} {Nat. Phys.}\ }\textbf {\bibinfo {volume} {3}},\
  \bibinfo {pages} {700} (\bibinfo {year} {2007})}\BibitemShut {NoStop}%
\bibitem [{\citenamefont {Loktev}\ and\ \citenamefont
  {Turkowski}(2009)}]{s.BKT.graphene.SC.PhysRevB.2009}%
  \BibitemOpen
  \bibfield  {author} {\bibinfo {author} {\bibfnamefont {V.~M.}\ \bibnamefont
  {Loktev}}\ and\ \bibinfo {author} {\bibfnamefont {V.}~\bibnamefont
  {Turkowski}},\ }\href {\doibase 10.1103/PhysRevB.79.233402} {\bibfield
  {journal} {\bibinfo  {journal} {Phys. Rev. B}\ }\textbf {\bibinfo {volume}
  {79}},\ \bibinfo {pages} {233402} (\bibinfo {year} {2009})}\BibitemShut
  {NoStop}%
\bibitem [{\citenamefont {Richter}\ \emph {et~al.}(2013)\citenamefont
  {Richter}, \citenamefont {Boschker}, \citenamefont {Dietsche}, \citenamefont
  {Fillis-Tsirakis}, \citenamefont {Jany}, \citenamefont {Loder}, \citenamefont
  {Kourkoutis}, \citenamefont {Muller}, \citenamefont {Kirtley}, \citenamefont
  {Schneider},\ and\ \citenamefont {Mannhart}}]{s.Richter.Nature2013}%
  \BibitemOpen
  \bibfield  {author} {\bibinfo {author} {\bibfnamefont {C.}~\bibnamefont
  {Richter}}, \bibinfo {author} {\bibfnamefont {H.}~\bibnamefont {Boschker}},
  \bibinfo {author} {\bibfnamefont {W.}~\bibnamefont {Dietsche}}, \bibinfo
  {author} {\bibfnamefont {E.}~\bibnamefont {Fillis-Tsirakis}}, \bibinfo
  {author} {\bibfnamefont {R.}~\bibnamefont {Jany}}, \bibinfo {author}
  {\bibfnamefont {F.}~\bibnamefont {Loder}}, \bibinfo {author} {\bibfnamefont
  {L.~F.}\ \bibnamefont {Kourkoutis}}, \bibinfo {author} {\bibfnamefont
  {D.~A.}\ \bibnamefont {Muller}}, \bibinfo {author} {\bibfnamefont {J.~R.}\
  \bibnamefont {Kirtley}}, \bibinfo {author} {\bibfnamefont {C.~W.}\
  \bibnamefont {Schneider}}, \ and\ \bibinfo {author} {\bibfnamefont
  {J.}~\bibnamefont {Mannhart}},\ }\href
  {http://dx.doi.org/10.1038/nature12494} {\bibfield  {journal} {\bibinfo
  {journal} {Nature}\ }\textbf {\bibinfo {volume} {502}},\ \bibinfo {pages}
  {528} (\bibinfo {year} {2013})}\BibitemShut {NoStop}%
\bibitem [{\citenamefont {Kessler}\ \emph {et~al.}(2010)\citenamefont
  {Kessler}, \citenamefont {Girit}, \citenamefont {Zettl},\ and\ \citenamefont
  {Bouchiat}}]{s.BKT.graphene.proximity.SC.PhysRevLett.2010}%
  \BibitemOpen
  \bibfield  {author} {\bibinfo {author} {\bibfnamefont {B.~M.}\ \bibnamefont
  {Kessler}}, \bibinfo {author} {\bibfnamefont {i.~m. c.~O.}\ \bibnamefont
  {Girit}}, \bibinfo {author} {\bibfnamefont {A.}~\bibnamefont {Zettl}}, \ and\
  \bibinfo {author} {\bibfnamefont {V.}~\bibnamefont {Bouchiat}},\ }\href
  {\doibase 10.1103/PhysRevLett.104.047001} {\bibfield  {journal} {\bibinfo
  {journal} {Phys. Rev. Lett.}\ }\textbf {\bibinfo {volume} {104}},\ \bibinfo
  {pages} {047001} (\bibinfo {year} {2010})}\BibitemShut {NoStop}%
\bibitem [{\citenamefont {Han}\ \emph {et~al.}(2014)\citenamefont {Han},
  \citenamefont {Allain}, \citenamefont {Arjmandi-Tash}, \citenamefont
  {Tikhonov}, \citenamefont {Feigelman}, \citenamefont {Sacepe},\ and\
  \citenamefont {Bouchiat}}]{s.BKT.graphene.proximity.SC.Nat.Phys.2014}%
  \BibitemOpen
  \bibfield  {author} {\bibinfo {author} {\bibfnamefont {Z.}~\bibnamefont
  {Han}}, \bibinfo {author} {\bibfnamefont {A.}~\bibnamefont {Allain}},
  \bibinfo {author} {\bibfnamefont {H.}~\bibnamefont {Arjmandi-Tash}}, \bibinfo
  {author} {\bibfnamefont {K.}~\bibnamefont {Tikhonov}}, \bibinfo {author}
  {\bibfnamefont {M.}~\bibnamefont {Feigelman}}, \bibinfo {author}
  {\bibfnamefont {B.}~\bibnamefont {Sacepe}}, \ and\ \bibinfo {author}
  {\bibfnamefont {V.}~\bibnamefont {Bouchiat}},\ }\href
  {http://dx.doi.org/10.1038/nphys2929} {\bibfield  {journal} {\bibinfo
  {journal} {Nat. Phys.}\ }\textbf {\bibinfo {volume} {10}},\ \bibinfo {pages}
  {380} (\bibinfo {year} {2014})}\BibitemShut {NoStop}%
\bibitem [{\citenamefont {Beasley}\ \emph {et~al.}(1979)\citenamefont
  {Beasley}, \citenamefont {Mooij},\ and\ \citenamefont
  {Orlando}}]{s.BKT.Beasley.PhysRevLett.1979}%
  \BibitemOpen
  \bibfield  {author} {\bibinfo {author} {\bibfnamefont {M.~R.}\ \bibnamefont
  {Beasley}}, \bibinfo {author} {\bibfnamefont {J.~E.}\ \bibnamefont {Mooij}},
  \ and\ \bibinfo {author} {\bibfnamefont {T.~P.}\ \bibnamefont {Orlando}},\
  }\href {\doibase 10.1103/PhysRevLett.42.1165} {\bibfield  {journal} {\bibinfo
   {journal} {Phys. Rev. Lett.}\ }\textbf {\bibinfo {volume} {42}},\ \bibinfo
  {pages} {1165} (\bibinfo {year} {1979})}\BibitemShut {NoStop}%
\bibitem [{\citenamefont {Homes}\ \emph {et~al.}(2004)\citenamefont {Homes},
  \citenamefont {Dordevic}, \citenamefont {Strongin}, \citenamefont {Bonn},
  \citenamefont {Liang}, \citenamefont {Hardy}, \citenamefont {Komiya},
  \citenamefont {Ando}, \citenamefont {Yu}, \citenamefont {Kaneko},
  \citenamefont {Zhao}, \citenamefont {Greven}, \citenamefont {Basov},\ and\
  \citenamefont {Timusk}}]{s.Homes.law.Nat.2004}%
  \BibitemOpen
  \bibfield  {author} {\bibinfo {author} {\bibfnamefont {C.~C.}\ \bibnamefont
  {Homes}}, \bibinfo {author} {\bibfnamefont {S.~V.}\ \bibnamefont {Dordevic}},
  \bibinfo {author} {\bibfnamefont {M.}~\bibnamefont {Strongin}}, \bibinfo
  {author} {\bibfnamefont {D.~A.}\ \bibnamefont {Bonn}}, \bibinfo {author}
  {\bibfnamefont {R.}~\bibnamefont {Liang}}, \bibinfo {author} {\bibfnamefont
  {W.~N.}\ \bibnamefont {Hardy}}, \bibinfo {author} {\bibfnamefont
  {S.}~\bibnamefont {Komiya}}, \bibinfo {author} {\bibfnamefont
  {Y.}~\bibnamefont {Ando}}, \bibinfo {author} {\bibfnamefont {G.}~\bibnamefont
  {Yu}}, \bibinfo {author} {\bibfnamefont {N.}~\bibnamefont {Kaneko}}, \bibinfo
  {author} {\bibfnamefont {X.}~\bibnamefont {Zhao}}, \bibinfo {author}
  {\bibfnamefont {M.}~\bibnamefont {Greven}}, \bibinfo {author} {\bibfnamefont
  {D.~N.}\ \bibnamefont {Basov}}, \ and\ \bibinfo {author} {\bibfnamefont
  {T.}~\bibnamefont {Timusk}},\ }\href {\doibase 10.1038/nature02673}
  {\bibfield  {journal} {\bibinfo  {journal} {Nature}\ }\textbf {\bibinfo
  {volume} {430}},\ \bibinfo {pages} {539} (\bibinfo {year}
  {2004})}\BibitemShut {NoStop}%
\bibitem [{s.p()}]{s.privcom}%
  \BibitemOpen
  \href@noop {} {}\bibinfo {note} {Private communications}\BibitemShut
  {NoStop}%
\end{thebibliography}%


\begin{thebibliography}{0}%
\makeatletter
\providecommand \@ifxundefined [1]{%
 \@ifx{#1\undefined}
}%
\providecommand \@ifnum [1]{%
 \ifnum #1\expandafter \@firstoftwo
 \else \expandafter \@secondoftwo
 \fi
}%
\providecommand \@ifx [1]{%
 \ifx #1\expandafter \@firstoftwo
 \else \expandafter \@secondoftwo
 \fi
}%
\providecommand \natexlab [1]{#1}%
\providecommand \enquote  [1]{``#1''}%
\providecommand \bibnamefont  [1]{#1}%
\providecommand \bibfnamefont [1]{#1}%
\providecommand \citenamefont [1]{#1}%
\providecommand \href@noop [0]{\@secondoftwo}%
\providecommand \href [0]{\begingroup \@sanitize@url \@href}%
\providecommand \@href[1]{\@@startlink{#1}\@@href}%
\providecommand \@@href[1]{\endgroup#1\@@endlink}%
\providecommand \@sanitize@url [0]{\catcode `\\12\catcode `\$12\catcode
  `\&12\catcode `\#12\catcode `\^12\catcode `\_12\catcode `\%12\relax}%
\providecommand \@@startlink[1]{}%
\providecommand \@@endlink[0]{}%
\providecommand \url  [0]{\begingroup\@sanitize@url \@url }%
\providecommand \@url [1]{\endgroup\@href {#1}{\urlprefix }}%
\providecommand \urlprefix  [0]{URL }%
\providecommand \Eprint [0]{\href }%
\providecommand \doibase [0]{http://dx.doi.org/}%
\providecommand \selectlanguage [0]{\@gobble}%
\providecommand \bibinfo  [0]{\@secondoftwo}%
\providecommand \bibfield  [0]{\@secondoftwo}%
\providecommand \translation [1]{[#1]}%
\providecommand \BibitemOpen [0]{}%
\providecommand \bibitemStop [0]{}%
\providecommand \bibitemNoStop [0]{.\EOS\space}%
\providecommand \EOS [0]{\spacefactor3000\relax}%
\providecommand \BibitemShut  [1]{\csname bibitem#1\endcsname}%
\let\auto@bib@innerbib\@empty
\end{thebibliography}%
\putbib[main_supp]
\end{bibunit}
\end{document}